\documentclass[preprint, aps, pre, eqsecnum,amsmath,amssymb,showpacs]{revtex4}

\usepackage[final]{graphicx} 

\newcommand{\bean}{\begin{eqnarray}}
\newcommand{\eean}{\end{eqnarray}}
\newcommand{\bea}{\begin{eqnarray*}}
\newcommand{\eea}{\end{eqnarray*}}
\newcommand{\beq}{\begin{equation}}
\newcommand{\eeq}{\end{equation}}

\def\lesssim{\mathrel{\mathpalette\vereq<}}
\def\vereq#1#2{\lower3pt\vbox{\baselineskip1.5pt \lineskip1.5pt
\ialign{$\hfill##\hfil$\crcr#2\crcr\sim\crcr}}}

\newcommand{\dd}{\text{d}}

\newcommand{\om}{\overline{m}}
\newcommand{\oalpha}{\overline{\alpha}}

\newcommand{\kb}{k_{\text{B}}}

\begin{document}
\renewcommand{\thefootnote}{\fnsymbol{footnote}}
\draft
\title{Magnetic properties of colloidal suspensions of interacting magnetic
particles}
\author{B.~Huke and M.~L\"{u}cke \\}
\affiliation{Institut f\"{u}r Theoretische Physik,Universit\"{a}t des Saarlandes,
D-66041 Saarbr\"{u}cken, Germany \\}

\renewcommand{\thefootnote}{\arabic{footnote}}
\setcounter{footnote}{0}
\date{\today}

\begin{abstract}
We review equilibrium thermodynamic properties of systems of magnetic particles
like ferrofluids in which dipolar interactions play an important role. The
review is focussed on two subjects: ({\em i}) the magnetization with the initial
magnetic susceptibility as a special case and ({\em ii}) the phase transition
behavior. Here the condensation ("gas/liquid") transition in the subsystem of
the suspended particles is treated as well as the isotropic/ferromagnetic
transition to a state with spontaneously generated long--range magnetic order.
\end{abstract}

\maketitle
\tableofcontents

\clearpage
\section{Introduction}

Ferrofluids \cite{R85} are suspensions of ferromagnetic or 
ferrimagnetic particles with diameters of the order of 10 nm in a liquid 
carrier. A common combination is magnetite (Fe$_3$O$_4$) in hydrocarbons.
The volume fraction of the magnetic material is 
typically a few percent. The particles are monodomain. They are too small
for the build--up of domain walls \cite{Ah00,Sp03}. Thus, the quantum--mechanical 
exchange interaction between the electronic spins can keep all the atomic 
magnetic moments aligned so that the total
magnetic moment, $m$, of the particles is of the order of about 10$^4$~$\mu_B$.
Hence, to describe the equilibrium magnetic properties of the ferrofluid one can
take each particle $i$ to have a magnetic
moment ${\bf m}_i (t)$ that fluctuates in time, however, with its magnitude
$m_i = |{\bf m}_i (t)|$ being fixed.
Thus, we do not address here the magnetic properties of the ferrofluid system 
on the nm scale, i.e., of the magnetic particles themselves:
Questions related to the size and form of the magnetocrystalline
anisotropy, to magnetostrictive effects, to shape anisotropy, to dead layers 
etc. of the particles \cite{Ah00,Sp03,DF92,Garcia00} are not addressed here. 

Whether the magnetic moment ${\bf m}_i (t)$ fluctuates only as a result of the
rotational diffusion of the particle in the carrier liquid (Brownian 
relaxation) or also because the temperature is sufficiently high to stimulate 
${\bf m}_i (t)$ itself to move in the potential landscape over the anisotropy
energy barriers of the particle's lattice structure (N{\'e}el relaxation) does not 
influence the
{\em long--time equilibrium} magnetic properties of the suspension but only its
dynamics. In any case the magnetic particles cause the ferrofluid to display the
equilibrium thermodynamics of a paramagnetic liquid --- if one postpones for the
moment the currently debated question of whether spontaneous 
long--range magnetic order is possible or not.  
The initial susceptibility can easily reach values of $\chi \approx 1$ and 
higher. Being 
orders of magnitude larger than for ordinary (molecular) paramagnetic 
liquids one thus speaks of a "superparamagnetic" liquid. 

Ferrofluid particles are small enough to avoid segregation caused by gravity or
customary magnetic gradient fields. But the strong van der Waals attraction
between the nanoscale particles would lead to irreversible agglomeration. To 
prevent this, 
they are either coated with polymer surfactants of about 2 nm length 
whose entropic repulsion counteracts the  van der Waals forces, or with 
ionic groups that prevent agglomeration due to their electrostatic repulsion. In
the latter case the carrier is a polar liquid like water, containing the
necessary counterions. 

In a sufficiently diluted ferrofluid the magnetic particles can be thought as
noninteracting, and the equilibrium magnetic properties of such a ferrofluid 
are those of an ideal paramagnetic 
gas. Still, there is an important difference to a molecular paramagnetic fluid
like, say, oxygen: the ferrofluid particles are not identical, they differ both
in
size and magnetic moment. This polydispersity has influence on the properties
of the ferrofluid, in particular also on the equilibrium magnetization.

At higher concentrations the ideal gas approximation fails to reproduce the 
equilibrium magnetization curve of ferrofluids. This is already the case for 
susceptibilities of the 
order of unity. Then, the particle interaction --- and first of all 
the dipolar interaction --- has to be taken into account. Since there exists a 
perfect analogy
between magnetic and electric dipoles, the large number of theories developed
for the latter case are directly applicable to this problem.

The thermodynamics of dipolar interacting particles is not only of interest
because of the applicability to (electric) polar fluids and ferrofluids. It is 
also of large theoretical interest because the dipole--dipole interaction shows
at least two
interesting features: ({\em i}) it is attractive and/or repulsive, depending on the
orientation of the particles and ({\em ii}) it is of long--range nature.

Because the dipolar interaction is neither always attractive nor always 
repulsive 
it is not trivially clear how the magnetic moments of the particles affect,
 e.~g., \ the
condensation phase transition of the suspended particles and especially whether 
the dipole--dipole
interaction can cause such a transition in a system of particles without
additional attractive interactions. This question is still being discussed
in the literature. An externally applied magnetic field can be expected to 
further modify the phase transition behavior.

Another example of a phase transition where the magnetic moments play an
important role is the isotropic/ferromagnetic transition, i.e., a spontaneous
breaking of the rotational symmetry resulting in a nonvanishing magnetization 
without applied magnetic field. Also this question whether the dipolar 
interaction alone can trigger such a phase transition is not yet settled.

There is quite an extensive literature on ferrofluids. The references to it 
are accumulated in 
the special issues of the Journal of Magnetism and Magnetic Materials that 
cover the
International Conferences on Magnetic Fluids. Odenbach et al. \cite{OTM02} 
show in the latest issue that by 2002 the total number of papers on
ferrofluids has grown to well above 6000 with about 1000 being published in the
period 1998--2001 \cite{MFB02}. The research topics 
cover problems as different as fluid dynamics and the modifications in classical
pattern formation experiments by using ferrofluids, pattern formation processes
genuine to ferrofluids as the Rosensweig instability or the labyrinth 
instability, magnetorheological properties,
dynamical magnetic susceptibility, medical and engineering applications
\cite{BMK93,BB96,BCM97,Odenbach02_LNP_m71,Odenbach02_LNP_594}.

We restrict ourselves here to the topics where the dipolar interaction between
the magnetic particles is of major importance, namely the already mentioned
equilibrium magnetization and initial susceptibility on the one hand, and phase
transitions on the other hand.

The review is organized as follows. Following this introduction we discuss
in Sec.~\ref{SecII} basic statistical mechanics of systems of dipolar particles. 
In Sec.~\ref{SecIII} the influence of the dipolar interaction
on the initial susceptibility of systems of otherwise spherical symmetric
particles is reviewed, mainly for dipolar hard spheres and Lennard--Jones
particles with an additional magnetic moment (Stockmayer particles). This 
question has already been discussed for the electrical case before ferrofluids 
became an active research area. The more general question of the equilibrium
magnetization on the other hand was mainly investigated with regard to the
application to ferrofluids. The literature concerning the equilibrium 
magnetization is reviewed in Sec.~\ref{SecIV}. In Sec.~\ref{SecV} the effect of
additional nonspherical interactions other than the dipolar one are 
briefly discussed. Phase transitions of the gas/liquid and
isotropic/ferromagnetic type are reviewed in Sec.~\ref{SecVI}. We conclude in
Sec.~\ref{SecVII}. 

Because of the analogy between systems of magnetically and electrically polar
particles and for reasons of simplicity, we will always use the magnetical
terminology when reviewing the literature, even when only the electric case is
considered. This is mostly the case in the older literature.

\clearpage

\section{Statistical mechanical description} \label{SecII}
A complete description of ferrofluids within the framework of statistical
mechanics would have to include all the components of a ferrofluid: magnetic 
particles, carrier liquid, polymer surfactants etc. But theoretical approaches 
to explain the 
thermodynamic properties connected to magnetism generally 
concentrate quite successfully on the magnetic component. Since the magnetic 
properties of the carrier liquid can be neglected, it is clear that it can only
have 
indirect influence on the magnetization. A more elaborate model of a ferrofluid 
describing both the magnetic particles and the carrier fluid is discussed by 
Kalikmanov \cite{K92}, who shows that under some assumptions the 
carrier fluid has no influence on the equilibrium magnetic properties of the 
ferrofluid as a whole. 
 
\begin{figure}
\includegraphics[width=16cm]{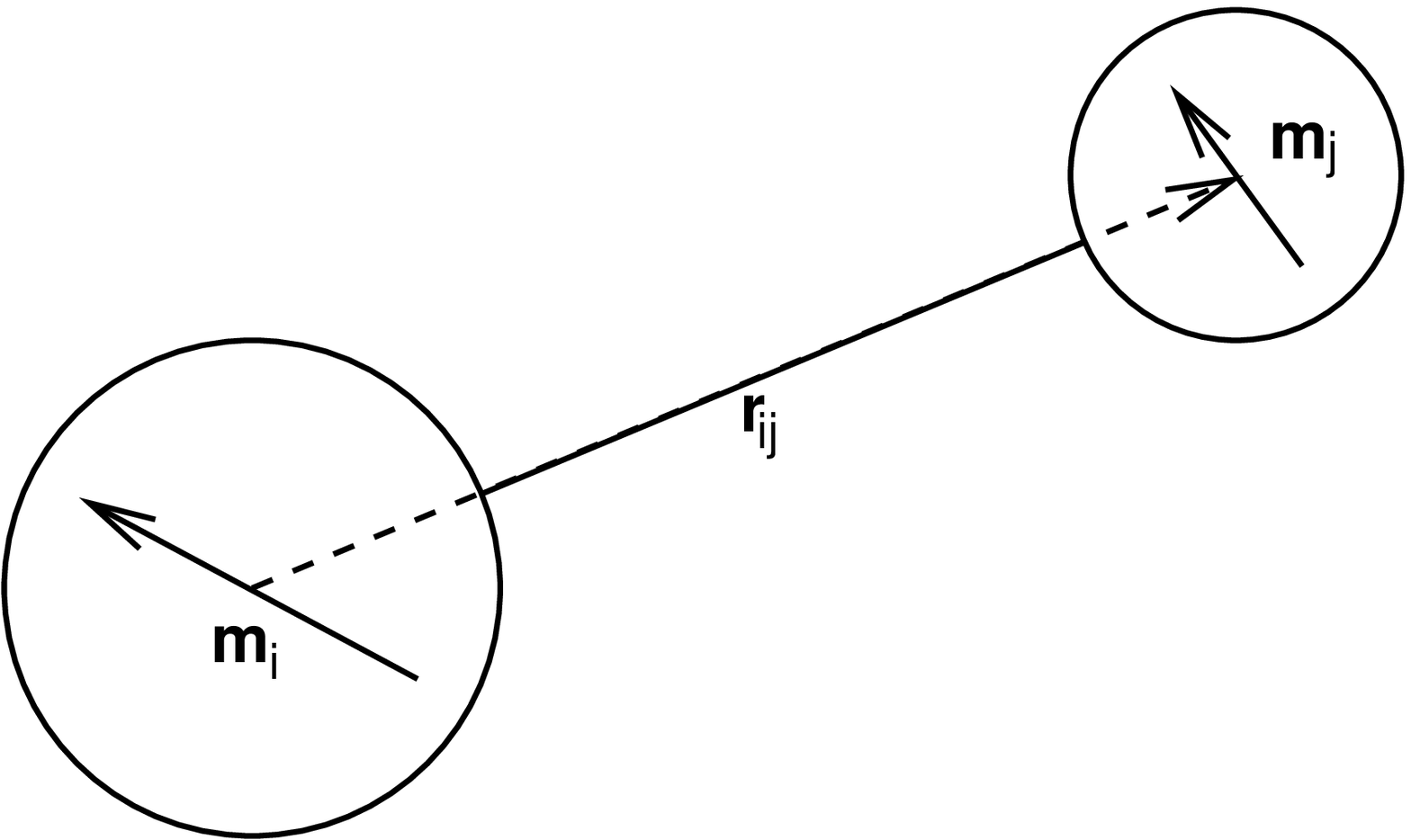}
\caption{Spheres $i$ and $j$ carrying dipole moments ${\bf m}_i$ and ${\bf m}_j$,
respectively, that interact via the potential (\ref{DHSVDDihdef}).}
\label{dipolar-spheres}
\end{figure}

Ignoring the carrier, the ferrofluid can be described as a system of 
particles, $i = 1 \, ... N$, each
carrying a magnetic moment ${\bf m}_i$ and thus interacting with an external
magnetic field ${\bf H}_e$ via the potential
\begin{equation}
V_{i} = - {\bf m}_i \cdot {\bf H}_e \;\; .
\end{equation}
The magnetic moments as used here include the vacuum permeability $\mu_0$, see e.g.
Eq.~(\ref{Eq:M_ideal_para}).
The absolute values $|{\bf m}_i|$ are fixed; but in general they are different 
from each
other. These magnetic particles interact pairwise with each other via the 
dipole--dipole interaction potential
\begin{equation}
V_{ij}^{DD} = - \frac{ 3 ({\bf m}_i \cdot {\bf \hat{r}}_{ij} )({\bf m}_j \cdot 
{\bf \hat{r}}_{ij} ) - {\bf m}_i \cdot {\bf m}_j }
{4 \pi \mu_0 r_{ij}^3} \;\; . \label{DHSVDDihdef}
\end{equation}
Here ${\bf r}_{ij} = r_{ij} {\bf \hat{r}}_{ij}$ is the distance vector between
particles $i$ and $j$ (see Fig.~\ref{dipolar-spheres}).
Since this potential decreases radially only as $r^{-3}$ the dipolar interaction
is of long--range nature. 

In addition to (\ref{DHSVDDihdef})
there generally exist also some short--range interactions between the particles 
that we capture by the potential $V_{ij}^{SR}$. It might include nonmagnetic 
components and in general also 
short--range magnetic interactions involving quadrupolar or higher order 
magnetic moments. For later use we define the dimensionless potentials
\begin{equation}
v_i = \frac {V_i}{\kb T} \; , \; v^{DD}_{ij} = \frac {V^{DD}_{ij}}{\kb T} \; , \; 
v_{ij}^{SR} = \frac {V_{ij}^{SR}}{\kb T} \, .
\end{equation}

The long--range nature of the magnetic interaction requires some care in 
statistical mechanical considerations. The local magnetic field acting on a
particle is the sum of the external field and the dipolar fields of the other
particles. The dipolar contribution depends on the distribution of the particles
on a macroscopic scale, i.~e.\ in particular also on the probe geometry.
In macroscopic magnetostatics this property translates
into the fact that the equilibrium magnetization ${\bf M}$ of a
magnetic medium depends on the internal field ${\bf H}$. This macroscopic
field depends itself on ${\bf H}_e$, ${\bf M}$, and 
geometry. The difference between ${\bf H}$ and ${\bf H}_e$ can therefore
be interpreted as an effect of dipolar interaction.

\subsection{Ideal paramagnetism}
\label{sec2a}
If all dipolar interactions are discarded then the magnetic particles of the 
ferrofluid feel only the external magnetic field, i.e., $H = H_e$. This is an 
useful approximation if $M \ll H_e$ such that $H 
\approx H_e$. Assuming in addition all particles to be identical  -- the 
monodisperse case -- with common magnetic moment $|{\bf m}_i| = m$, then
the equilibrium magnetization is given by
\begin{equation} \label{Eq:M_ideal_para}
M = M_L = \frac{m N}{\mu_0 V} {\cal L} \left( \frac{m H}{\kb T} \right) =
M_{sat}{\cal L} \left( \alpha \right) \; 
\end{equation}
with $H=H_e$. Here $N/V$ is the particle density and $\cal L$ is the Langevin 
function,
${\cal L}(\alpha) = \coth(\alpha) -1/\alpha$. Its argument $\alpha = mH/\kb T$ 
measures the energy of the moment $m$ in the field $H=H_e$ in units of the 
thermal energy $\kb T$. The saturation magnetization $M_{sat}=m N/\mu_0 V$ of 
the ferrofluid corresponds to the case of parallel dipoles. 

One might think that using in Eq.~(\ref{Eq:M_ideal_para}) the real internal 
field $H$ in the ferrofluid instead of $H_e$ would be an improvement of the
ideal paramagnetic model by itself. However such a replacement does not 
incorporate the effect of dipolar interactions in a systematic and controlled 
manner since they additionally modify the simple functional relation 
(\ref{Eq:M_ideal_para}) between equilibrium magnetization $M$ and internal 
field $H$ in various ways that are of the same order as the replacement of 
$H_e$ by $H$ in Eq.~(\ref{Eq:M_ideal_para}).

In a polydisperse extension of the above described ideal, noninteracting 
monodisperse model the magnetic moments of the particles are allowed to be 
different. Then the magnetization is given by
\begin{equation}
M = M_L = \sum_i \frac{m_i}{\mu_0 V} 
{\cal L} \left( \frac{m_i H}{\kb T} \right)  \;\; .\label{Meqiu_poly}
\end{equation}
Here the sum extends over all magnetic particles, $i=1, ...,N$.
Using the mean magnetic moment $\om = (1/N) \sum m_i$ to introduce the reduced
moment $\mu_i = m_i/\om$ of particle $i$ we can write more conveniently
\begin{equation}
M = \frac{\om}{\mu_0 V} \sum_i \mu_i {\cal L} 
\left( \frac{\mu_i \om H}{\kb T} \right) =
M_{sat}{\cal L}^{\text{poly}} \left( \oalpha \right) \;\; ,
\end{equation}
where
\begin{equation} \label{EQ:L_poly}
M_{sat}=\frac{\om N}{\mu_0 V} \;\; , \qquad 
{\cal L}^{\text{poly}} \left( \oalpha \right) = \frac{1}{N}
\sum_i \mu_i {\cal L} 
\left( \mu_i \oalpha \right)\;\; , \qquad \oalpha = \frac{\om H}{\kb T} \;\; .
\end{equation}

\subsection{Weiss and Onsager model}
\label{secweiss}
The earliest model of a selfinteracting magnetic medium is the mean-field 
Weiss model \cite{Debye12}. In this monodisperse model every particle is 
thought to be located in the center of an empty spherical cavity that is 
surrounded by a magnetic continuum with an
internal field $H$ and magnetization $M$. In thermal equilibrium,
the magnetization is given by a Langevin function into which enters the 
local field $H_{local} = H + M/3$ within the spherical cavity. This leads to 
the implicit relation
\begin{equation}
M = M_{sat} {\cal L} \left[ \frac{m}{\kb T} \left(H + \frac{M}{3} \right) \right]\;\; ,
\label{Weissformula}
\end{equation}
that can be solved numerically to give the 
sought after equilibrium magnetization $M(H)$. 

The initial susceptibility of the Weiss model
\begin{equation}
\chi  = \frac{\partial M}{\partial H} ( H = 0 ) = 
\frac{\chi_L}{1 - \chi_L/3} \;\; , \label{WOchiweiss}
\end{equation}
is equivalent to the well known Clausius--Mosotti relation. Here
\begin{equation}
\chi_L  = \frac{\partial M_L}{\partial H} ( H = 0 ) = 
\frac{1}{3} M_{sat} \frac{m}{\kb T} 
\end{equation}
is the
Langevin initial susceptibility of the ideal monodisperse paramagnetism 
(\ref{Eq:M_ideal_para}).

The Weiss model works well for weakly interacting ferrofluids but strongly 
overestimates the magnetization of concentrated magnetic fluids. It predicts
ferromagnetic solutions, i.~e., a spontaneous magnetization $M(H=0)\neq 0$
whenever $\chi_L > 3$. While this criterion is at odds with experiments 
the possible existence of ferromagnetic phases in systems of dipolar hard spheres 
cannot be ruled out for stronger interactions 
(c.f. \ Sec.\ \ref{SEC:ferrophase}).

A similar early approach to the problem of a selfinteracting magnetic medium
is the Onsager theory \cite{O36} originally conceived for polarizable molecules.
Therein, say, spherical molecules occupy a cavity in polarizable continuum with
susceptibility $\chi$. The field acting on the molecule is the sum of a cavity
field plus a reaction field that is parallel to the actual total (permanent 
and induced) moment of the molecule. The resulting equation 
between $\chi_L$ and $\chi$ reads here
\begin{equation}
\frac{\chi(3 + 2 \chi)}{3(1 + \chi)} = \chi_L \label{WOchionsager} \;\; .
\end{equation}

While the Weiss model overestimates the initial susceptibility of
concentrated ferrofluids, the Onsager model generally underestimates it.
It is instructive to write $\chi$ according to (\ref{WOchiweiss}) and 
(\ref{WOchionsager}) for small $\chi_L$ as power series in $\chi_L$
\begin{equation}
\chi = \chi_L + \frac{1}{3} \chi_L^2 \pm \frac{1}{9} \chi_L^3 + ...\;\; .
\end{equation}
Here the positive sign refers to the Weiss model and the negative one to the 
Onsager model. The first and second term agree with a systematic expansion of
$\chi$ in terms of dipolar coupling strength and magnetic volume fraction 
(c.f. Sec.~\ref{SEC:clus-exp-meth}). However, the third order 
term of this expansion reads $+\frac{1}{144} \chi_L^3$ (see Eq.~\ref{CLEoep2}). 
Thus both, Onsager as well as Weiss model strongly overestimate the influence 
of this term. 
\subsection{Short--range potentials}
\label{SEC:short--range-pot}
If one wants to go beyond the mean field models of Sec.~\ref{secweiss} one has
to specify the potential with which the magnetic particles interact 
besides the long--range dipolar one. 

 Rosensweig \cite{R85} has proposed a quite
elaborate potential for the short--range interaction, $V^{SR}$, of sterically 
stabilized magnetic particles. It consists of the sum of the
van der Waals attraction between spherical magnetic particles and the sterical 
repulsion of their surfactants, i.e., $V^{SR} = V^{\text{vdW}} + V^{\text{ster}}$.
The van der Waals attraction between the spheres is given by
\begin{equation}
V^{\text{vdW}}_{ij}(r_{ij}) = - \frac{A}{6} \left[
\frac{2}{l^2 + 4l} + \frac{2}{(l + 2)^2} + \ln \frac{l^2 + 4l}{(l + 2)^2} 
\right] \;\; . \label{vdWpot}
\end{equation}
Here $A$ is the Hamaker constant. The quantity 
\begin{equation}
l = \frac{2s}{D_{\text{core}}} = \frac{2 r_{ij}}{D_{\text{core}}} -2 \;\; 
\label{SYlabstanddef}
\end{equation}
is given by twice the surface-to-surface distance $s$ of the particles 
reduced by the diameter $D_{\text{core}}$ of the magnetic particles while 
$r_{ij}$ is the distance between their centers.
The sterical repulsion of the surfactants is described by
\begin{equation} \label{stericalpot}
V^{\text{ster}}_{ij}(r_{ij}) = \frac{\pi D_{\text{core}}^2 \xi \kb T}{2} 
\left[
2 - \frac{l+2}{t} \ln \left( \frac{1+t}{1+l/2} \right) - \frac{l}{t}
\right] \qquad 0 < l < 2 t \;\; .
 \end{equation} 
Here $\xi$ is the surface density of the polymers and 
\begin{equation}
t = \frac{2 \delta_{\text{polymer}}}{D_{\text{core}}} 
\end{equation}
is given by twice the length, $\delta_{\text{polymer}}$, of the surfactant 
polymers. In the Rosensweig potential the carrier liquid component
is not completely ignored but rather taken into account
on a continuum level via the electric permeability of the liquid that enters 
into the Hamaker constant $A$. 

\begin{figure}
\includegraphics[width=10cm,angle=270]{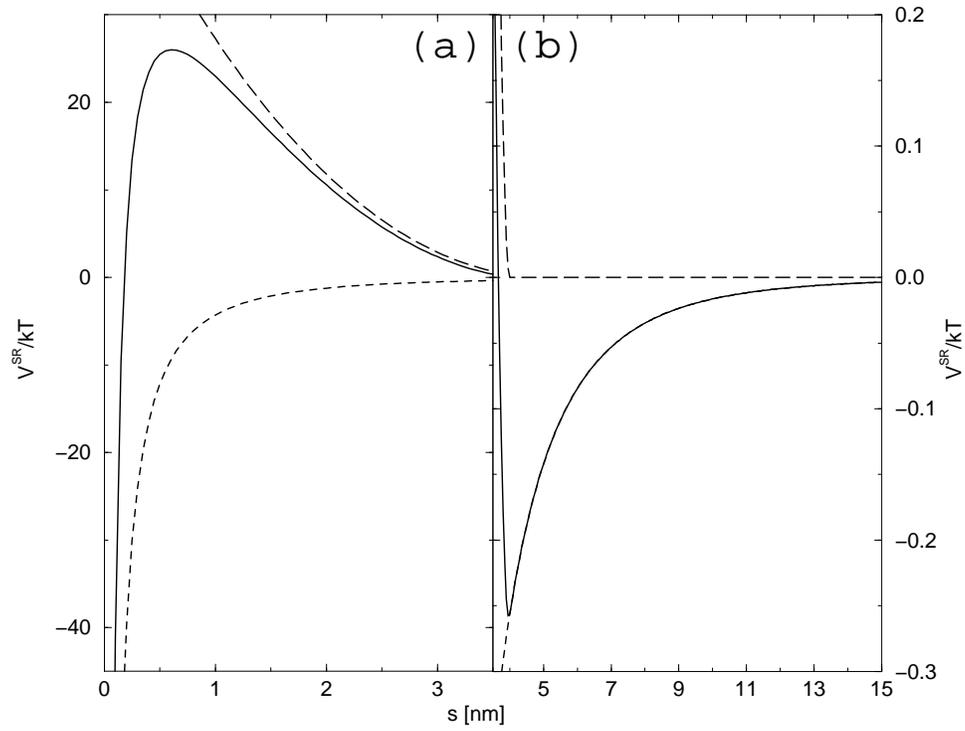}
\caption{Non--magnetic interaction potential for sterically stabilized
ferrofluid particles as function of the surface-to-surface distance $s$. Short 
dashed 
line: van der Waals attraction, long dashed line: steric repulsion, solid 
line: combined potential. Note that the scales of both separation and 
energy are different in (a) and (b).}
\label{rosensweig1}
\end{figure}

Fig.~\ref{rosensweig1} shows the interaction potential 
$V^{SR} = V^{\text{vdW}} + V^{\text{ster}}$ (\ref{vdWpot}, \ref{stericalpot}) 
for typical values 
given by Rosensweig: $A = 10^{-19} \, J$, $\xi = 1 \, \text{nm}^{-2}$, 
$D_{\text{core}} = 10 \, \text{nm}$, and $\delta_{\text{polymer}} = 2 \, 
\text{nm}$. The need for sterical
stabilization becomes apparent in this plot. The van der Waals interaction is
strongly attractive for small particle separations, it even diverges in this
simple model. The sterical potential modifies the interaction for distances less
than 4 nm, such that only a small attractive tail at larger distances remains
in the combined potential. At very small surface-to-surface separations the
potential still diverges, but this inner attractive region is effectively
shielded by a potential barrier with a height of about 25 $\kb T$. 
 
To obtain an equivalent model potential for electrostatically
stabilized ferrofluids, $V^{\text{ster}}$ is replaced (see e.~g.\ \cite{KMR87}) by
the interaction potential 
\begin{equation}
V_{ij}^{\text{Debye}}(r_{ij}) = \frac{Q_{\text{eff}}^2}{4\pi \epsilon r_{ij}}
e^{-( r_{ij} - D_{\text{core}})/\lambda_d}  
\label{SYdebyedef}
\end{equation}
of charged spheres in ionic solutions. Here $\lambda_d$ is the Debye length, 
$\epsilon$ is the dielectric constant of the carrier liquid, and 
$Q_{\text{eff}}$ is an effective charge of the particles. The combined potential
shows qualitatively the same features as in the sterical case.

Another important model potential for describing dipolar particles is the
Stockmayer potential in which the short--range interaction
is given by the Lennard-Jones function for the 
van der Waals interaction of point-like particles
\begin{equation} \label{LJPotential}
v^{SR}_{ij} = 4 v_0 \left[ \left( \frac{D}{r_{ij}} \right)^{12} - 
\left( \frac{D}{r_{ij}} \right)^{6} \right] \;\; .
\end{equation}
Here $D$ is the collision parameter marked by the zero of
this potential and $v_0$ is the depth of the potential well.

However, most often the effects of particle interaction are discussed within 
the simpler model of dipolar hard spheres (DHS). Therein the short--range 
part of the interaction is given by the hard sphere repulsion potential
\begin{equation}
v^{SR}_{ij} = v^{HC}_{ij} = \left\{ \begin{array}{cl}
\infty & \qquad \mbox{for $r_{ij} < D$} \\
0 & \qquad \mbox{for $r_{ij} > D$}
\end{array} \right. \;\; , \label{DHSVHCijdef}
\end{equation}
with $D$ being the common hard sphere diameter in the monodisperse case.
In the polydisperse case with different hard sphere diameters, $D_i$, one has
to replace $D$ by $(D_i + D_j)/2$ in the above equation. The only other
interaction in this model is the long--range dipolar potential. 

Some aspects of the sterical interaction (\ref{stericalpot}) can also be
incorporated into the DHS model. Since the magnetic field of a perfect, 
homogeneously magnetized sphere is exactly
dipolar one can introduce into the DHS model particles that consist of 
two concentric spheres: an outer hard sphere with diameter $D$ 
and a magnetized spherical core of diameter $D_{\text{core}} < D$. The latter is 
the source of the dipole field. The nonmagnetic layer models the surfactants,
i.e., $D - D_{\text{core}} \approx 2 \delta_{\text{polymer}}$. A nonmagnetic
dead layer \cite{SISI87} on the particle can also be modeled.

\begin{figure}
\includegraphics[width=10cm,angle=270]{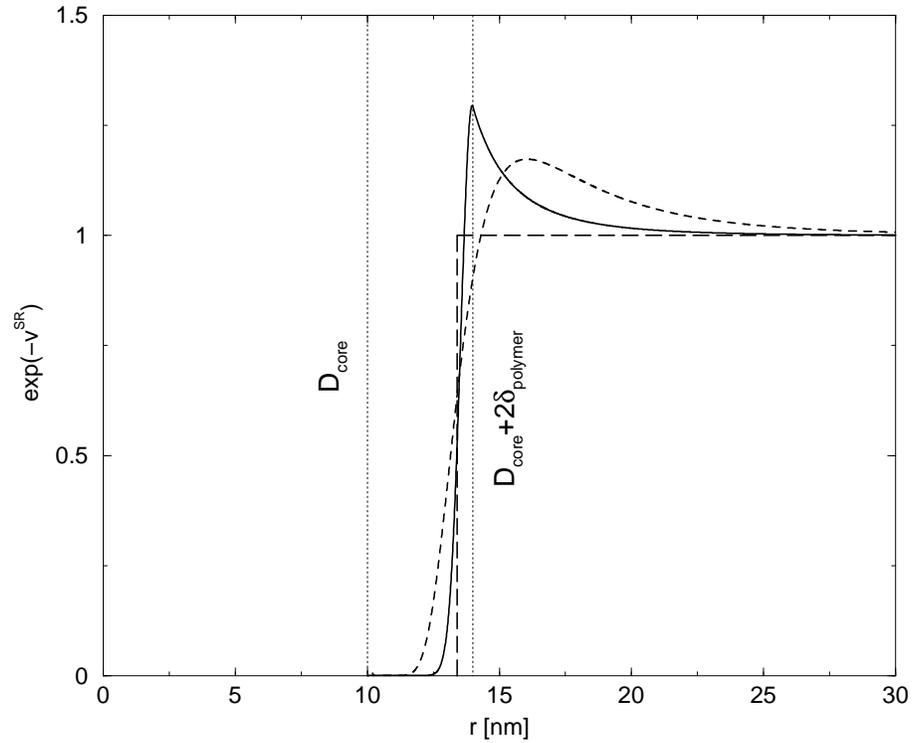}
\caption{The quantity $\exp(-v^{SR})$ as function of particle distance for the
potential of Fig.~\ref{rosensweig1} (solid line). The contribution from the 
divergent part of the potential at very small surface-to-surface distances is 
suppressed in this plot. The short
dashed and long dashed lines show $\exp(-v^{SR}(r))$ for best fitting hard 
sphere and van der Waals potentials, respectively.}
\label{rosensweig2}
\end{figure}

The DHS model system was initially introduced to describe the electric 
polarizability of fluids of polar molecules. However, in reality deviations 
from the spherical shape, particle polarizability, and higher order moments 
have an important influence. Ferrofluids, on the other hand, resemble much
better the model system. 
That the hard sphere repulsion is a good approximation to the Rosensweig potential
becomes apparent when looking at the quantity $\exp(-v^{SR})$ which is more 
relevant for the equilibrium properties than $v^{SR}$ itself. In 
Fig.~\ref{rosensweig2} $\exp[-v^{SR}(r)]$ is plotted for the Rosensweig 
potential (solid line) ignoring the irrelevant inner attractive region.
As can be seen, $\exp(-v^{SR})$ resembles quite well the unit step function 
$\exp(-v^{HC})$ of the hard sphere potential that goes from 0 to 1 at contact 
distance.

The long dashed line for the best fitting
hard sphere potential and the short dashed line for the van der Waals potential 
are included in Fig.~\ref{rosensweig2} for comparison. Although the Rosensweig 
potential resembles more the
van der Waals potential since both have an attractive part, the hard sphere
potential fits the Rosensweig potential even a little bit better.
\clearpage
\section{Initial susceptibility} \label{SecIII}
Most of the older approaches to the magnetic or electric properties of polar
fluids
deal with the linear response problem, i.~e.\ the initial susceptibility. The 
nonlinear behavior was of lesser interest in common atomic polar fluids 
since available electromagnetic fields were not strong enough to reach
saturation. However, with the advent of superparamagnetic ferrofluids with 
large magnetic
moments of the particles the linear and also the nonlinear magnetization
behavior of interacting dipoles became again a more active research area.

Most of the approaches to calculate the initial susceptibility can roughly be 
classified into cluster expansion, hypernetted chain, and numerical simulation 
methods. In this section, we will mainly address the results for 
dipolar hard spheres.
\subsection{Cluster expansion methods}
\label{SEC:clus-exp-meth}
Consider a system of $N$ particles interacting with each other via a pair 
potential $v_{ij}$ and with an external potential $v_i$. 
Then the canonical partition function is given by 
\begin{equation}
Z = \int \exp \left( -\sum_i  v_i 
- \sum_{i<j} v_{ij} \right) \; \dd \Gamma \;\; . 
\label{CLEpartfunc}
\end{equation} 
Here $d \Gamma$ indicates the integration over the configuration space spanned
by the positions and magnetic moments of the particles. The ideal-gas factor 
contained in the canonical partition function that comes from the momentum 
degrees of freedom is not indicated here.

The key point of
the cluster expansion method is to introduce into (\ref{CLEpartfunc}) the
functions
\begin{equation}
f_{ij} = \exp \left(- v_{ij} \right) -1 \;\; 
\label{CLEfterm}
\end{equation} 
which are small when the interaction is weak. Thus, one inserts 
$\exp \left(- v_{ij} \right) = 1+f_{ij}$ into (\ref{CLEpartfunc}) and expands
the integrand in terms of $f_{ij}$. 
 
Then (\ref{CLEpartfunc}) reads up to first order for example 
\begin{equation}
Z = Z^0_N + Z^0_{N-2} \sum_{i<j} \int f_{ij} \; \dd (i) \dd (j) \;\; , 
\label{CLEpartfunc1}
\end{equation} 
where $Z^0_N$ is the partition function of the ideal noninteracting 
$N$-particle system. These approximations to $Z$ then allow 
to calculate the thermodynamic quantities of interest. The advantage of this 
expansion is that in contrast to (\ref{CLEpartfunc}) the approximations to $Z$ 
such as (\ref{CLEpartfunc1}) and higher orders require
only low dimensional integrations, that can be performed at least numerically.

In the case of DHS 
the calculation of the leading terms can be done even analytically if one 
performs another series expansion, namely the expansion of (\ref{CLEfterm})
in terms of the dipolar interaction potential $v_{ij}^{DD}$ that enters via
$v_{ij} = v_{ij}^{DD} + v_{ij}^{HC}$. The two expansions then translate into a 
power 
series for $Z$ and thus for $M$ or $\chi$ in two dimensionless parameters,
namely the volume fraction
\begin{equation}
\phi = \frac{N \pi D^3}{6 V}
\end{equation}
of the particles, and a dipolar coupling constant
\begin{equation}
\lambda = \frac{m^2}{4 \pi \mu_0 D^3 \kb T} \;\; .
\end{equation}
It is also common to choose $\rho^* = N D^3/V$ and $y = 8 \phi \lambda /3$
as parameters. These parameters fully characterize the system for $H = 0$.
To make a comparison with Stockmayer particles easier, we will use these
parameters here also. In the case of Stockmayer particles one also has to specify 
the potential strength $v_0$ entering into Eq.~(\ref{LJPotential}).
It should be noted that in the literature a system of 
Stockmayer particles is normally characterized by giving values for $1/v_0$ and
$\lambda/v_0$.  

In ordinary ferrofluids, $\phi$ is of the order of 10$^{-2}$. For
magnetite ferrofluids with 10 nm particle diameters and an additional polymer 
layer of 2 nm $\lambda$ is less than one. In this case, higher orders in the 
$(\phi,\lambda)$--expansion have only
minor influence on $\chi$. But it is also possible to produce
much stronger interacting ferrofluids. For example, van Ewijk, Vroege, and 
Philipse \cite{EVP02} report the production of ferrofluids with values for 
$\lambda$ up to 2.7 and volume fractions up to 0.5. Even higher $\lambda$ could
be realized by Mamiya, Nakatani, and Furubayashi \cite{MNF00}.

The long--range nature of the dipolar interaction causes some cluster
integrals to be geometry dependent and mathematically ambiguous if the
thermodynamic limit is naively performed. This problem is circumvented by
dealing with a macroscopic but finite spherical geometry. In that case 
the initial susceptibility with respect to the external field
$\chi_e = \partial M / \partial H_e \, (H_e = 0)$ can be calculated unambiguously
and is related to the geometry--independent quantity $\chi$ via 
\begin{equation}
\frac{\chi}{3 + \chi } = \frac{1}{3} \chi_e \;\; . 
\end{equation}

In 1966 Jepsen \cite{J66} performs a cluster expansion to obtain
\begin{equation}
\frac{\chi}{3 + \chi } = \frac{1}{3} \chi_L - \frac{5}{144} \chi_L^3 
\label{CLEjespen}
\end{equation}
for DHS. $\chi$ is then up to the calculated order given by 
\begin{equation}
\chi =  \chi_L + \frac{1}{3} \chi_L^2 + \frac{1}{144} \chi_L^3 \;\; .
\label{CLEoep2}
\end{equation}
This equation can also be understood as a $(\phi, \lambda)$--expansion, since
$\chi_L = M_{sat}m/ 3 \kb T =  N m^2/3 \mu_0 \kb T V$ can be written as 
\begin{equation}
\chi_L = 8 \phi \lambda \,. 
\end{equation}
The fact that both the Weiss and the Onsager model
correctly predict the prefactor of the $\chi_L^2$--term shows that this term 
has a mean-field 
origin being independent of the type of short--range interaction.

Rushbrooke \cite{R79} calculates more terms in 1979 . His result, as corrected 
later by Joslin \cite{J81}, reads
\begin{eqnarray}
\frac{\chi}{3 + \chi } &=& \frac{8}{3} \phi \lambda
+ \frac{64}{225} \phi^2 \lambda^4
- \frac{160}{9} \phi^3 \lambda^3   
- \frac{8(1187-600 \ln 2)}{3375} \phi^3  \lambda^4 + 1.13358 \; \phi^4 
\lambda^3 
\nonumber \\
&=& 2.66667 \phi \lambda + 0.284444 \phi^2 \lambda^4 - 
17.7778 \phi^3 \lambda^3 \\ && \qquad \nonumber
-1.82782\phi^3  \lambda^4 + 1.13358 \; \phi^4 \lambda^3\;\; .
\end{eqnarray}
The $\phi \lambda$-- and $\phi^3 \lambda^3$--terms agree with the $\chi_L$--
and $\chi^3_L$--terms of Jepsen. The last term requires a numerical 
integration. Rushbrooke´s error affected the $\phi^3  \lambda^4$--term.
Buckingham and Joslin \cite{BJ80} calculate the second dielectric virial
coefficient, i.~e.\ the 
$\phi^2 \lambda^n$--terms in the above expansion:
\begin{equation}
\frac{\chi}{3 + \chi } = ... \; + 
\sum_{n=1}^\infty \frac{64}{3 n \left[(2n + 3)!! \right]^2} 
\sum_{k=0}^n (3k -n) \left( \begin{array}{c} 2k \\ k \end{array} \right)
\phi^2 \lambda^{2n+2} 
+ \; ... \;\; . \label{CLEBJ80}
\end{equation}
Joslin \cite{J81} presents besides the aforementioned correction to the 
result of Rushbrooke also the term
\begin{equation} 
\frac{32(315 \ln 2 - 218)}{3375} \phi^3 \lambda^5 =
0.00323662 \phi^3 \lambda^5 \;\; .
\end{equation}
Thereafter Rushbrooke and Shrubsall \cite{RS85} calculate in addition the term
\begin{equation} 
\left[ \frac{105133619}{4042500} - \frac{298976}{7875} \ln 2 \right] 
\phi^3 \lambda^6 = -0.308396 \phi^3 \lambda^6 \;\; .
\end{equation}
A numerical calculation of the $\phi^3$--term as a function of $\lambda$
using Monte--Carlo methods is reported by Joslin and Goldman in 1993 \cite{JG93}.

Tani~et~al.~\cite{THB83} take a slightly different road
to calculate $\chi$ using the system of (nonpolar) hard spheres as reference
system. They get
\begin{equation} 
\chi = 8 \phi \lambda + \frac{64}{3} \phi^2 \lambda^2 + f(\phi) \lambda^2 
\;\; , \label{CLETanifdef}
\end{equation} 
where the last term is a numerical expression based on approximate expressions
for the two-- and three--particle correlation functions for hard spheres as
found in Monte--Carlo simulations.  For small $\phi$ this result reduces to
(\ref{CLEoep2}). Goldman \cite{G90} compares results of this theory very
successfully to Monte--Carlo data for systems of hard spheres and Stockmayer 
particles.

\subsection{Hypernetted chain and related approaches}
Using the hypernetted chain (HNC) approach the initial susceptibility is 
calculated via the two--particle correlation functions $g(1,2)$ and 
$h(1,2) = g(1,2) - 1$ of the polar particles \cite{HMcD90}. The HNC is 
based on the Ornstein--Zernike relation
\begin{equation}
h(1, 2) =  c(1,2) + \frac{N}{4 \pi V} \int c(1,3) h(3,2) \dd (3)
\;\; , \label{HNCOZdef}
\end{equation}
defining the direct correlation function $c(1,2)$ and a closure relation, the 
HNC approximation reading
\begin{mathletters}
\begin{eqnarray}
c(1,2) = h(1,2) - \ln[g(1,2)] - v_{12} \label{HNCHNCgleich} &\qquad
r_{12} > D \;\; , \\
g(1,2) = 0 & \qquad r_{12} < D
\end{eqnarray}
\end{mathletters}
in the case of hard spheres.
It can be shown \cite{HMcD90} that this approximation is equivalent to a cluster
expansion using an infinite number of $f$--integrals, neglecting only a certain
class, the so--called bridge diagrams that appear only in higher 
orders. 

Wertheim \cite{W71} and Nienhuis and Deutch \cite{ND72} used the so--called mean 
spherical
approximation (MSA) \cite{LP66} to calculate the initial susceptibility of DHS.
In the MSA (\ref{HNCHNCgleich}) is replaced by 
\begin{equation}\label{Eq:MSA}
c(1,2) = - v_{12} \qquad \qquad \text{for }  r_{12} > D \;\; ,
\end{equation}
effectively using the approximation 
$\ln[g(1,2)] = \ln[1 + h(1,2)] \approx h(1,2)$, which is correct for large
distances, where $h(1,2)$ is small. 
Wertheim showed that this equation together with (\ref{HNCOZdef}) allows a simple
ansatz, representing the dependence of the correlation functions on  
${\bf m}_1$ and ${\bf m}_2$ as linear combinations of three functions. He sets
\begin{equation}
h(1,2) = h_S (r_{12}) + h_D(r_{12}) f_D(\hat{\bf m}_1,\hat{\bf m}_2) +
h_\Delta(r_{12}) f_\Delta(\hat{\bf m}_1,\hat{\bf m}_2) 
\label{M7MSAhansatz}
\end{equation}
with 
\begin{mathletters}
\begin{eqnarray}
f_\Delta(\hat{\bf m}_1,\hat{\bf m}_2) &=& \hat{\bf m}_1 \cdot \hat{\bf m}_2 
\;\; , \\
f_D(\hat{\bf m}_1,\hat{\bf m}_2) &=& 
3(\hat{\bf r}_{12} \cdot \hat{\bf m}_1) 
(\hat{\bf r}_{12} \cdot \hat{\bf m}_2) -
\hat{\bf m}_1 \cdot \hat{\bf m}_2 \;\; .
\end{eqnarray}
\end{mathletters}
$g(1,2)$ and $c(1,2)$ have a similar representation with the same 
$f_D, f_\Delta$.
The MSA then provides an implicit solution for $\chi$ that is given by the
three equations
\begin{equation}
\chi = \frac{\chi_L}{q(-x)} \;\; , \qquad
\chi_L = q(2x)-q(-x) \;\; , \qquad
q(x) = \frac{(1 + 2x)^2}{(1-x)^4} \;\; . \label{HNCMSAgleich}
\end{equation} 
Here, $\chi$ depends only on $\chi_L = 8 \phi \lambda$ but not on $\phi$ and 
$\lambda$ separately. Expanding $\chi$ in powers of $\chi_L$ yields again
(\ref{CLEoep2}) for the leading terms.

Verlet and Weis \cite{VW74} propose in 1974 an improved theory. Therein they
replace the 
term $h_S(r)$ that does not depend on the dipolar character of the fluid 
by a better result for nonpolar hard spheres. Stell and Weis 
\cite{SW77} calculate the initial susceptibility of DHS using this modification
and a further improved version. They get larger values for $\chi$ that agree
better with the Monte--Carlo data.

Patey \cite{P77} improves the theory of Wertheim by retaining the ansatz
(\ref{M7MSAhansatz}) but expanding the full relation (\ref{HNCHNCgleich}) 
linear in $f_D$ and $f_\Delta$. This approach is known as linear HNC (LHNC),
although it was pointed out later that an equivalent theory was proposed
already in 1973 by Wertheim \cite{W73} himself, called  single--superchain 
theory. In contrast to the MSA, the LHNC requires a numerical 
calculation of the correlation functions  $h_S(r)$, $h_D(r)$, etc.\
to find the initial susceptibility. Within the LHNC $\chi$ depends on both
$\phi$ and $\lambda$ independently. Patey did not apply the "pure" HNC but
uses again $h_S$ and $c_S$ from the system of nonpolar hard spheres as system of
reference (known as reference LHNC or RLHNC). 

The RLHNC was followed by the reference quadratic HNC (RQHNC) proposed by
Patey, Levesque and Weis. In \cite{PLW79} this approach
is applied to DHS. In the RQHNC (\ref{HNCHNCgleich}) is expanded up to quadratic
terms in $f_D$ and $f_\Delta$. Differences in $\chi$ between RQHNC and RLHNC 
amount only up to a few 
percent for $\lambda < 2$ and relatively large $\phi \approx 0.4$. But they are
more significant for larger $\lambda$. However, later on it turned out 
in Monte--Carlo simulations (see Sec.~\ref{numcal}) that both theories
overestimate $\chi$ here --- the apparently better RQHNC even more so than the 
RLHNC.

These authors considered also Stockmayer particles, using DHS as a reference 
system.
They argue that a Stockmayer system with $\phi = 0.8 \, \pi/6$ and 
$1/v_0 = 1.35$ should be similar to a system of DHS with the same $\phi$
for not too large $\lambda$. Indeed, the results for the susceptibility do not
differ very much for $\lambda < 2$. For higher coupling constants the RQHNC 
now gives smaller values for $\chi$ than the RLHNC.

Agrafonov, Martinov and Sarkisov \cite{AMS80} expand the full
HNC equation in terms of $\lambda$ and obtain the result
\begin{equation}
\frac{\chi^2}{\chi + 1} = 64 \phi^2 \lambda^2 \left( 1 - \frac{8}{3} \phi
\lambda \right)  
\end{equation}
for DHS. This expression yields after solving for $\chi$ the correct 
terms in $\phi \lambda$ and
$\phi^2 \lambda^2$. However, the next term is already incorrect.
Chan and Walker \cite{CW82} propose to expand the HNC 
approximation (\ref{HNCHNCgleich}) given in the form
\begin{equation}
g(1,2) = \exp[ h(1,2) - c(1,2) - v_{12} ] \;\; , 
\end{equation}
in terms of rotational invariants for the orientational distribution 
of $\hat{\bf m}_1$ and $\hat{\bf m}_2$, truncating the expansion after 
a sufficient number of terms. They take into account however only three terms,
namely the functions already used in the  MSA, RLHNC, and RQHNC. Using this
truncated HNC ansatz (THNC) they obtain even higher values for the
susceptibility of DHS than predicted by the RQHNC.

In 1985 Fries and Patey \cite{FP85} finally extend the ansatz
(\ref{M7MSAhansatz}) by adding more terms. They also use 
(\ref{HNCHNCgleich}) in a form differentiated with respect to $r_{12}$ 
eliminating the logarithmic term. Solving numerically the full 
RHNC approximation for DHS in that way they come to results for $\chi$ that are 
in much better 
agreement with Monte--Carlo calculations. In the same year Lee, Fries, and Patey 
\cite{LFP85} also investigate the Stockmayer system. As in the case of DHS
the results agree better with numerical data than those of RLHNC and RQHNC.
Finally, in 1986, 
Fries and Patey \cite{FP86} apply the similar Percus--Yevick approximation to
DHS finding it less well suited to predict the susceptibility than the RHNC and 
even the MSA.

Another modification is done by Lado \cite{Lado86},
who varies the diameter of the nonpolar hard spheres in the system of 
reference to minimize the free energy.
Lomba, Martin and Lombardero \cite{LML92} solve the pure
HNC finding for the correlation functions even better agreement with 
Monte--Carlo calculations than for the RHNC.

\subsection{Numerical simulations}
\label{numcal}

Numerical simulations for calculating the susceptibility are mostly based on
Monte--Carlo methods. $\chi$ is either determined from the magnetization 
fluctuations  $\left< M^2 \right>$ in the absence of an external field or
more directly from the magnetization in small fields. Calculating the
susceptibility from the simulation results for $h(1,2)$ are less appropriate 
since $\chi$ depends on the long--range behavior of $h(1,2)$ that cannot be
determined accurately in finite simulation cells \cite{PLW79}.

The influence of the distant
dipoles can be incorporated by cutting off the dipolar interaction at a finite
distance $R_s$ and replacing the dipoles beyond $R_s$ by a magnetic continuum
of given $\chi$ (reaction field method). This susceptibility is either 
adjusted to the susceptibility of the simulated system or set to an arbitrary 
value (commonly to $\chi = 0$). The chosen value then enters into the 
determination of the system's susceptibility. 

Another method that is the most commonly used today is the Ewald summation based
on a periodic 
continuation of the system together with an effective method for calculating the
resulting total dipolar field. A discussion of the different methods can
be found e.~g.\ in \cite{A80,AA81}. The Ewald summation is 
described in \cite{AMcD76}.

Monte--Carlo calculations of the free energy and other thermodynamical
quantities that depend less strongly on the long--range nature of the dipolar
forces were already performed in the seventies of the last century 
\cite{BW73,PV73,PV74,VW74}. Patey et~al.~\cite{P77,PLW79,LPW77} calculate
the correlation function $h(1,2)$  and compare  with the predictions of the
LHNC and QHNC. An indirect comparison for the susceptibility, demonstrating 
the superiority of the MSA over the Onsager model can be found already in
\cite{PV74}. 

An early Monte--Carlo calculation of $\chi$ is done by Adams and McDonald \cite{AMcD76}
in 1976. The authors however do not consider a fluid but f.~c.~c.\ and s.~c.\
lattices. They compare Ewald sum and reaction field results for interaction
strengths up to $\phi \lambda \approx 2$. Levesque, Patey, and Weis 
\cite{LPW77} calculate $\chi$ for $\phi = 0.8 \pi/6$ and $\lambda
\leq 1$. They compare different methods and system sizes and in 1982 they also
study the case $\lambda = 2$ \cite{PLW81}. Adams \cite{A80} considers already 
in 1980 
$\lambda = 2.75$. Lado \cite{L86} uses a combination of Monte--Carlo 
calculation and RHNC for the long--range part of the potential. 

Susceptibility calculations for Stockmayer particles were mainly performed 
for the values $\phi = 0.8 \pi/6$ and $v_0 = 1/1.35$ as already considered by 
Patey, Levesque and Weis \cite{PLW79}. Pollock and Alder \cite{PA80} consider 
values for $\lambda$ up to 3 while Adams and Adams \cite{AA81} study 
different values for $v_0$. Levesque and Weis 
\cite{LW84} calculate $\chi$ for $\lambda = 2$, and Evans and Morriss 
\cite{EM85} do so for $\lambda \leq 1.7$. Neumann, Steinhauser, and Pawley
\cite{NSP84} consider the parameter combination $v_0=1/1.15$, 
$\phi \approx 0.43$,  and $\lambda = 2.6$. Hesse--Bezot, Bossis, and 
Brot \cite{HBB84} use in 1984 molecular dynamic methods to investigate 
$\lambda = 2$.

Some of the works cited above treat the long--range nature of the dipolar 
interaction incorrectly, see the discussion by Neumann \cite{Neu83},
Neumann and Steinhauser \cite{Neu80,NS85}, and Gray et~al.~\cite{GSJ86}. 
These papers also discuss the relationship between the reaction field and the
Ewald sum approach. Further discussions of the Ewald 
summation and of other techniques used for dipolar systems were given recently 
by Wang, Holm, and M{\"u}ller \cite{WH01,WHM_JCP03}.

\subsection{A comparison}
\label{comparison}
\begin{figure}
\includegraphics[width=10cm,angle=270]{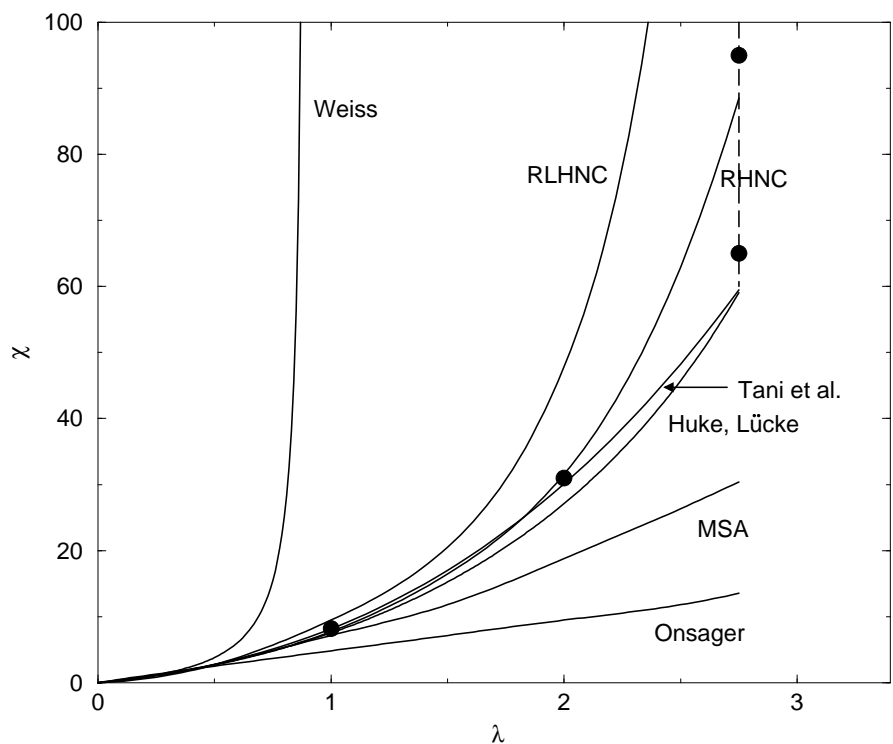}
\caption
{Comparison of theoretical results for the initial susceptibility of dipolar
hard spheres for $\phi = 0.8 \pi/6$: Weiss \cite{Debye12} and 
Onsager \cite{O36} models,
MSA by Wertheim \cite{W71}, RLHNC results by Patey \cite{P77}, RHNC results
by Fries and Patey \cite{FP85}, cluster expansion result by Huke and
L{\"u}cke \cite{BHML2000}, and the theory by Tani~et~al.~\cite{THB83}. 
Full circles denote Monte--Carlo data taken from a similar figure in 
\cite{HMcD90}.}
\label{CompareHMcDo}
\end{figure}

Unfortunately most of the older works deal with systems with large volume
fractions: a typical value is $\phi = 0.8 \pi/6 = 0.419$. This is reasonable 
for customary polar liquids but it is very high for ferrofluids. 
Fig.~\ref{CompareHMcDo} shows a comparison of the most
important theories discussed above for the case of DHS. 
In this figure $\chi$ is plotted as a function of
$\lambda$ for fixed $\phi = 0.419$. The Monte--Carlo data 
(denoted by circles) and the results for RLHNC and RHNC were taken from 
\cite{FP85} and a similar plot in \cite{HMcD90}. Onsager theory and MSA
clearly underestimate the Monte--Carlo data while Weiss theory and RLHNC 
overestimate them.
The RHNC works much better, as does the numerically much less complicated
result by Tani et al.\ \cite{THB83}. The curve denoted as "Huke and L{\"u}cke" 
refers to a cluster expansion theory \cite{BHML2000} for the full magnetization 
curve. In the linear case considered here in this section this theory 
\cite{BHML2000} yields Eq.~(\ref{CLEoep2}) and in addition also the 
$\phi^2 \lambda^n$ terms given by Eq.~(\ref{CLEBJ80}). 

The large scatter of the Monte--Carlo results for 
$\lambda = 2.75$ is indicated in Fig.~\ref{CompareHMcDo} as well. Surprisingly
there seems to be no final conclusion 
in the literature concerning the appropriate value of $\chi$ for 
$\lambda = 2.75$ despite the fact that these different Monte--Carlo results 
have been in the literature for quite a while.
The different predictions for $\chi$ at $\lambda = 2.75$ are discussed 
in \cite{NS85} where preference is given to a relatively low value of 
$\chi \approx 64$.

\subsection{Other potentials}
The susceptibility of some variants of the DHS and Stockmayer particle systems 
were also investigated. The system of dipolar sticky hard spheres, where the
short--range potential contains an additional $\delta$--function term, were
investigated by Chapela and Martina \cite{CM85} using integral
theories like MSA and RLHNC to calculate $\chi$. The authors come to
contradictory results concerning the effect of the additional term 
in the potential on $\chi$.
Joslin and Gray \cite{JG86} show that the second dielectric virial
coefficient of dipolar sticky hard spheres is positive.

Kusalik \cite{Ku89,K90,Ku91} investigates in 1989 and the following years 
dipolar soft spheres, i.~e.\ Stockmayer particles without the attractive 
$r^{-6}$--term in the short--range part of the potential using the RLHNC and 
the RHNC as well as Monte--Carlo calculations. The Monte--Carlo results for 
$v_0 = 1/1.35$, $\phi = 0.8 \pi/6$, and $\lambda = 2$ are very similar to
those for Stockmayer particles. But RLHNC and RHNC overestimate $\chi$ somewhat
more.

Henderson, Boda, Szalai, and Chan \cite{HBSC99,SHBS99} apply the MSA, the theory
by Tani et~al.~\cite{THB83}, and Monte--Carlo calculations to dipolar Yukawa 
particles with a hard core. Again, the results are found to be similar to those
for dipolar hard spheres and Stockmayer particles. The Monte--Carlo data for $\chi$ at
$\phi = 0.8 \pi/6$ and $\lambda = 1,\, 2$ depend little on the strength of the 
Yukawa potential and they agree well with the Monte--Carlo data for dipolar hard spheres. The 
perturbation theory by Tani et~al. is found to reproduce the Monte--Carlo data better than
the MSA.

To summarize, the nature of the short--range interaction seems to have a rather
weak influence on the susceptibility, even for relatively high densities.

\subsection{Polydisperse theories} 

In real ferrofluids the influence of polydispersity can normally not be
neglected. In the case of ideal paramagnetism discussed in Sec.~\ref{sec2a}
the initial susceptibility is given by
\begin{equation}
\chi_L = \sum_i \frac{m_i^2}{3 \mu_0 V \kb T} \;\; ,
\end{equation}
as follows from (\ref{Meqiu_poly}). Therefore $\chi_L$ is proportional to the 
second moment of the distribution of the magnetic moments and larger than in the
monodisperse case if the average particle volume is kept constant. When particle
interactions are taken into account even higher moments of the distribution enter
into the initial susceptibility.

Several of the theories discussed above were also extended to the system of
polydisperse or at least bidisperse DHS, where the particles have different
hard sphere diameters and/or carry different magnetic moments. Already in 
1973 the MSA of Wertheim is applied to the polydisperse case by    
Adelman and Deutch \cite{AD73}, although restricted to particles with
a common diameter. The authors define an equivalent monodisperse system
having the same thermodynamical properties.   
 
Isbister and Bearman \cite{IB74} generalize the MSA to mixtures with 
arbitrary diameters. The susceptibility predicted by this theory is calculated 
by Freasier, Hamer, and Isbister \cite{FHI79}. The authors also
describe a comparable monodisperse system giving at least approximately the same
results. Ramshaw and Hamer simplify the evaluation \cite{RH81}.
Cummings and Blum \cite{CB86} compare results of the MSA 
to Monte--Carlo results for the bidisperse case.
 
Lee and Ladanyi present \cite{LL87} a RLHNC extension 
and generalizations \cite{LL89}  of the RHNC and the cluster expansion
by Tani et~al.\ \cite{THB83}. They compare to the Monte--Carlo data from
\cite{CB86}, coming to similar conclusions about the quality of the different
theories as in the monodisperse case discussed above (Sec.~\ref{comparison}). 

\subsection{Applications to ferrofluids}

With the preparation of highly concentrated ferrofluids real
physical systems became available that have much more in common with the simple model system of
DHS than the ordinary polar fluids. Ferrofluids thus allowed not only more 
reliable
experimental tests but they also initiated new theoretical investigations.  

Two--dimensional Monte--Carlo simulations on the initial susceptibility 
of ferrofluids are performed by  Menear, O'Grady, Bradbury et~al.~in the early 
eighties \cite{MGBCMPC83,MBC83,MBC84,BMC85,BMC86}. Here the particles
are not modeled as pure hard spheres but the sterical repulsion of the
surfactants according to Rosensweig (\ref{stericalpot}) is also taken into account.
In \cite{BMC86} a polydisperse ferrofluid is simulated. In particular, the 
temperature dependence of the susceptibility entering via 
$\lambda \sim 1/T$ is investigated. A Curie--Weiss law  
$1/\chi \sim T -T_0$ is found as to be expected from simple mean field models,
e.~g.\ the Weiss model.

Bradbury, Martin, and Chantrell \cite{BMC87,MBC87} also
perform full three--dimensional simulations for DHS with additional sterical
repulsion and obtain good agreement with RHNC calculations. The authors
again find a Curie--Weiss law for $\chi(T)$ but less pronounced than for the
2D calculations. The calculations are done for a small $\phi = 0.01$ and very
high $\lambda$ up to $7.21$. These values refer to the real hard sphere
diameter. However, the dipolar coupling would be better characterized by using
the total diameter of the particles, i.e., of core plus polymer layer. That 
would reduce $\lambda$ to values of maximal 2.63. In this $\lambda$-range 
much simpler cluster expansion theories still work and reproduce
the result. It should be mentioned that although 
$1/\chi = c( T - T_0)$ is a very good  approximation in that range, the Weiss
model already fails since it predicts significantly different values for $c$ 
and $T_0$. 

A behavior of $1/\chi \sim  T - T_0$ is also found in experiments
\cite{MGBCMPC83,PAA88}. The results in \cite{MGBCMPC83} however rely 
on data in a narrow range of temperatures of 60 K where only very strongly
interacting ferrofluids should exhibit a pronounced nonlinear behavior.
In \cite{PAA88} a negative value for $T_0$ is found, i.~e., a N{\'e}el
behavior that was not predicted by the theories. 
A more comprehensive experimental investigation of the initial susceptibility 
is published 
in 1990 by Holmes, O'Grady, and Popplewell \cite{HOGP90}. Inspecting
a broader range of temperatures the authors show deviations from the Curie--Weiss
law for ferrofluids based on magnetite. They also argue that the N{\'e}el 
behavior found in \cite{PAA88} may result from
measurements for too large fields, i.~e.\ already in the nonlinear range of
the magnetization curve. They find a linear and a quadratic term in the
concentration dependence of the susceptibility. The nonlinear 
behavior of $1/\chi(T)$ is confirmed by Williams et~al. \cite{WOCD93}
in 1993.

Morozov et~al. \cite{MPRS87,PLM87} apply the MSA to ferrofluids.
Further results can be found in \cite{SPMS90}. The authors demonstrate good
agreement with experiments concerning the concentration and temperature 
dependence of $\chi$. 
As discussed above in Sec.~\ref{comparison}, the MSA does not
seem to be a good theory for strongly interacting systems. The investigated
ferrofluids ($\phi \leq 0.172$, $\chi \leq 10$) however fall in range where
the difference in $\chi$ between MSA and cluster expansions \cite{THB83,BHML2000}
is only a few percent.
Applying the formula (\ref{CLEoep2}) shows that the $\chi_L^3$--term which is
correctly predicted by the MSA gives only minor contributions to
$\chi$.

Pshenichnikov \cite{P95} compares several theories, namely 
the Onsager model, the Weiss model, the MSA and a theory by Ivanov et~al.\ 
\cite{BI92} with experimental data in a similar range of $\phi$ and $\lambda$.
The latter theory \cite{BI92} deals with the equilibrium magnetization for 
arbitrary fields. It is explained in the next
section. For the initial susceptibility this theory gives 
$\chi = \chi_L + \chi_L^2/3$, i.~e., Eq.~(\ref{CLEoep2}) up to second order. 
Pshenichnikov shows good agreement with
both Ivanov's theory and the MSA, thereby also indirectly demonstrating the 
small influence of the $\chi_L^3$--term that is present in the MSA but not
in Ivanov's theory. Both theories however fail to adequately describe 
the temperature dependence of ferrofluids with $\chi(T)$ lying 
in the range between $\approx$ 20 and 80. 
 
Kalikmanov \cite{K99} proposes in 1999 the so-called algebraic
perturbation theory that is corrected by 
Szalai, Chan and Henderson \cite{SCH99} and found to be identical
to the cluster expansion theory of Tani et~al.~\cite{THB83}. 

Van Ewijk, Vroege, and Philipse \cite{EVP02} perform susceptibility 
measurements on highly concentrated ferrofluids and compare with different 
theories. For a ferrofluid with $\lambda \approx 2.7$ and $\chi_L \leq 6$, they 
find a surprisingly low susceptibility of $\chi \leq 10$, best described by the 
Onsager theory or the MSA. These results 
seem to be in contradiction to the data presented in Fig.~\ref{CompareHMcDo}, 
although $\phi$ is slightly higher there. Some theoretical support to these 
findings can however be found in a paper by Pshenichnikov and Mekhonoshin 
\cite{PM00a}, who present in 2000 Monte Carlo data in the same range of 
$\chi_L$. For $\lambda = 3$ and $\lambda =4$, they find similar low values for 
the susceptibility. They explain the low susceptibility with aggregates of
particles, that have a small combined magnetic moment and interact with the
field only weakly.
\newpage
\section{Magnetization}\label{SecIV}

The equilibrium magnetization of a
ferrofluid cannot be described in general by a simple Langevin ansatz or its
polydisperse generalization. Experimental hints to this fact can already be 
found in a paper from 1979 by Tari~et~al.~\cite{TCCP78} where a comparison of a 
magnetization curve with a theoretical curve based on an independently measured 
size distribution is made. However, the 
theories for the magnetization of, say, dipolar hard spheres in arbitrary
fields are less advanced than for the special case of the initial susceptibility.
Most of the theories discussed below have been devised with regard to the
application to ferrofluids. For this reason, many of them were directly or 
shortly after extended to include also the effects of polydispersity. 

When comparing a
polydisperse ferrofluid with a monodisperse one with the same average magnetic 
moment of the particles and the same volume fraction, the equilibrium 
magnetization curve will be steeper for small fields in the polydisperse case
since the initial susceptibility is higher. The asymptotic behavior for high
fields will however be the same. This results in a maximal difference between
the magnetization curves for medium fields $\alpha \approx 1$. 

Pollock and Alder \cite{PA80} and Adams and Adams \cite{AA81} perform Monte--Carlo 
calculations for the equilibrium magnetization of Stockmayer particles. 
In 1981 H{\o}ye and Stell \cite{HS81} propose a mean field model 
based on general thermodynamic considerations and compare their results with
the data from \cite{PA80}. The magnetization in this model
is given by
\begin{mathletters}
\begin{equation}
M = M_{sat} {\cal L} \left( \frac{m}{\kb T} H_{\text{eff}} \right) \;\; ,
\label{M7HSterm1}
\end{equation}
where $H_{\text{eff}}$ is an effective field
\begin{equation}
H_{\text{eff}} = H + \frac{1}{3} (1 - \Theta) M \;\; .
\label{M7HSterm2}
\end{equation}
\end{mathletters}
Here $\Theta$ is to be taken from the results for the initial susceptibility. 
For $\Theta =0$ the formula reduces to that of the Weiss theory.
Dikanskii \cite{Di82} proposes a concentration dependent term for an
effective field calculated from a fit to susceptibility data.
Sano and Doi~\cite{SD83} investigate phase transitions
in ferrofluids using a model of a randomly filled cubic lattice. As an 
additional result they obtain the Weiss expression for the equilibrium
magnetization.

Woodward and Nordholm \cite{WN84} propose an effective
potential for the dipolar interaction that results from averaging over the
orientations of the dipoles according to
\begin{equation}
\exp \left( - V^{\text{eff}}_{12} / \kb T \right) = 
\left< \exp \left( - V^{DD}_{12} / \kb T \right) 
\right>_{\hat{\bf m}_1, \hat{\bf m}_2} \; \; . \label{SYVeffdef}
\end{equation} 
The so defined $V^{\text{eff}}$ is then a function of particle separation and 
temperature.
It vanishes for large distances as $r_{12}^{-6}$. In 1986 \cite{WN86} these 
authors extend their ansatz to a theory for the equilibrium 
magnetization. They propose a functional for the free energy $F$
of a magnetic continuum in a magnetic field. Minimized
with respect to the orientational distribution of the dipoles it gives for zero
field an expression for $F$ that contains $V^{\text{eff}}$ as interaction 
potential.
Including the external field the dipolar interaction is described  by a long- 
range part as in the Weiss model and a short--range part that is a generalization 
of $V^{\text{eff}}$ which, however, depends also on the magnetic field.
The magnetization has the form (\ref{M7HSterm1},\ref{M7HSterm2}) with
a function $\Theta(M,T)$ that can be calculated numerically. The authors compare
the resulting
initial susceptibility with the RLHNC predictions.
The results are however not convincing: They are larger than in the RLHNC
already for $\phi \approx 0.3$ and diverge for $\phi \approx 0.4$. 
Comparisons for DHS with the magnetization results of H{\o}ye and Stell can 
be found in \cite{WN88}.

Berkovsky, Kalikmanov, and Filinov \cite{BKF85,BKF87,K92}
develop between 1985 and 1992 a thermodynamical theory for ferrofluids.
Based on cluster expansion methods they calculate the
magnetization in two special cases. They derive expressions
for spherical geometries where $H = H_e - M/3$.
For small fields $\alpha \ll 1$  the authors obtain for the susceptibility with
respect to the external field $H_e$ 
\begin{equation}
\chi_e = \frac{\partial M}{\partial H_e} = 8 \lambda \phi - 
\frac{160}{3} \lambda^3 \phi^3 \;\; . \label{SYchiedef}
\end{equation}
If one restates this expression into a result for $\chi$ a missing 
$\phi^2 \lambda^2$--term appears and one obtains (\ref{CLEoep2}). 
For stronger fields they get
\begin{equation}
M = M_{\text{sat}} \left[ {\cal L} (\alpha_e) + 
L_{1,2}(\alpha_e) \phi \lambda^2 f(\phi) \right] \;\; . \label{M7O12}
\end{equation}
Here $L_{1,2}(\alpha_e)$ is an analytic expression and
$f(\phi)$ depends on the correlation function $g(1,2)$ of nonpolar hard
spheres that is used as reference system. Again, by restating the expression in
a form $M = M(H)$ an additional term of order $\phi \lambda$ shows up.
Here one has to note that $M/M_{sat}$ is independent of $\phi$ and 
$\lambda$ in leading order, whereas $\chi$ starts with a $O(\phi
\lambda)$--term. Likewise, terms of order $\phi^m \lambda^n$ in $M/M_{sat}$
always correspond to $O(\phi^{m+1} \lambda^{n+1})$--terms in $\chi$.

Between 1990 and 1992 also Ivanov et~al.~\cite{BBI89,I89,BIZ90,I92,BI92} 
develop a thermodynamical theory of ferrofluids also using
cluster expansion techniques. For a needle--shaped geometry in which $H_e = H$
they obtain the correct result in $O(\phi \lambda)$
\begin{equation}
M = M_{\text{sat}} \left[ {\cal L}(\alpha) + 
8 \phi \lambda {\cal L}(\alpha) {\cal L}^\prime(\alpha)
\right] \;\; . \label{M7O11}
\end{equation}
A correct polydisperse generalization, replacing ${\cal L}(\alpha)$ by
${\cal L}^{\text{poly}}(\oalpha)$ (\ref{EQ:L_poly}), is also given.

Lebedev~\cite{Le89} suggests to apply the MSA also to finite magnetic 
fields. To that end he replaces the equations (\ref{HNCMSAgleich}) by
\begin{equation}
M(H) = \frac{M_L(H)}{q(-x)} \;\; , \qquad
\frac{\dd M(H)}{\dd H} = \frac{q(2x)-q(-x)}{q(-x)} \;\; , \label{M7MSAgleichmodify}
\end{equation} 
with $q(x)$ given as before by the third equation in (\ref{HNCMSAgleich}). Thus,
$\chi$ and $\chi_L$ in the first equation of (\ref{HNCMSAgleich}) is replaced in 
(\ref{M7MSAgleichmodify}) by $M$ and $M_L$, respectively. The second equation of 
(\ref{M7MSAgleichmodify}) can be derived by combining the first two equations
in (\ref{HNCMSAgleich}) and replacing the initial susceptibility 
$\chi$ by the susceptibility $\dd M(H)/\dd H$ for finite fields. Hence,
the equations (\ref{M7MSAgleichmodify}) reduce to the MSA for small fields.

Morozov et~al.~\cite{ML90,SPMS90} present in 1990 a real extension of the
MSA aimed at predicting the magnetization for arbitrary fields. They use the
so-called Lovett-Mou-Buff-Gubbins
equation \cite{LMB76,Gu80} to relate the one--particle probability density
$\rho({\bf r}, {\bf m})$ in the presence of an external field to the direct 
correlation function $c(1,2)$ taken in the 
MSA form of Eq.~(\ref{Eq:MSA}). With Wertheim's method \cite{W71} they then
derive from the Ornstein-Zernike equation two independent Percus-Yevick like
equations for a hard-sphere fluid with renormalized densities. As a result of
this approximation scheme they present $M$ given by the
Langevin magnetization for an effective field $C \alpha$
\begin{equation}
M =  M_{\text{sat}} {\cal L} \left( C \alpha \right) \;\; . \label{M7WeissmitHc}
\end{equation}
Here $C$ has to be calculated as follows: Let
\begin{mathletters}
\begin{equation}
A = \frac{\partial \ln M}{\partial \ln (C H)} \qquad \text{and} \qquad
35 x^2 - 5(10 - 3 A) x - 7 (7 + 3 A) = 0 \;\; . \label{SYadef}
\end{equation}
With $x_1$ and $x_2$ being two solutions of the quadratic equations, $C$ must be
given selfconsistently by
\begin{equation}\label{SYydef}
C = \left[ 1 + \frac{4 \pi}{3} ( 1 + x_1)\frac{M}{H} \right] \frac{1}{q(x_1 y)} = 
\left[ 1 + \frac{4 \pi}{3} ( 1 + x_2)\frac{M}{H} \right] \frac{1}{q(x_2 y)} \; .
\end{equation}
\end{mathletters}
Here $y$ is defined via the second equality and $q(x)$ is still given by 
the expression in (\ref{HNCMSAgleich}). The authors also propose a polydispersive
generalization. Comparisons with experiments can be found
in \cite{SPMS90,ML90}: The MSA reproduces well the magnetization curves of
polydisperse ferrofluids with saturation magnetizations up to 87~kA/m. 

Based on preliminary work \cite{MB85} concerning the 2D case  
Bradbury and Martin \cite{BM93} and independently 
Ayoub et~al.~\cite{ASADOL93} develop the so--called 
dimer model of dipolar interaction. The partition function for $N$ particles
is written as the $(N/2)$th power of a two-particle partition function calculated 
via a cluster expansion assuming that the two particles are closer than a 
typical next--neighbor distance. This model has several shortcomings. The
long--range character of the forces is not taken into account leading to
an expression for $M$ in which the $\phi \lambda$--term is missing. Also the 
calculated $O(\phi \lambda^2)$--terms are wrong in both cases and moreover
they do not agree with each other \cite{error-BM93-ASADOL93}. Even the 
corrected calculation
leads to a wrong result by a constant factor due to the somewhat arbitrary
assumption of using the next--neighbor distance as a cutoff.

Zubarev and Iskakova \cite{ZI96} present a magnetization equation relying 
on a model of noninteracting magnetic clusters (chains). They derive the
equilibrium distribution of chain lengths based on a simple expression for the
free energy of a system of such chains. The magnetization is then formally the 
equilibrium magnetization of a polydisperse ideal paramagnetic gas where the
chains are the fundamental constituents. The chain length distribution is however 
a function of the magnetic field and furthermore of $\phi$ and $\lambda$. 
Abu--Aljarayesh and Migdadi \cite{AM99} investigate this theory further,
calculating the entropy and some equilibrium magnetization curves.

Pshenichnikov, Mekhonoshin and Lebedev \cite{PML96} suggest an equation
for the equilibrium magnetization that reads
\begin{equation}
M = M_{\text{sat}} {\cal L} \left[ \alpha + \frac{m M_{\text{sat}}}{3 \kb T} 
{\cal L} ( \alpha ) \right ] \; .\label{M7O11korr}
\end{equation}
It agrees better with their experiments in the range of medium fields than
the expression (\ref{M7O11}) that gives the same initial susceptibility.
It furthermore gives more consistent results for the diameter distribution of
polydisperse ferrofluids at different stages of dilution. Furthermore, their
earlier result (\ref{M7O11}) can be derived from Eq.~(\ref{M7O11korr}) by 
expanding the
latter up to first order in $\phi$ and $\lambda$ since 
$m M_{\text{sat}}/3 \kb T = 8 \phi \lambda$. On the other hand, 
Eq.~(\ref{M7O11korr}) can be obtained from the Weiss formula (\ref{Weissformula})
by iterating and stopping after the first step. However, Eq.~(\ref{M7O11korr}) 
has the advantage over Eq.~(\ref{M7O11}) that it always produces physically sound
magnetization curves with $M$ being a monotonic function of $H$ with $0 \leq M <
M_{sat}$. This is not the case for Eq.~(\ref{M7O11}) when $\phi \lambda$ becomes 
too large. 

In 2000 the authors of this review \cite{BHML2000} perform a cluster expansion 
calculation to derive some of the terms in the sum
\begin{equation}
M = M_{sat} \left[ {\cal L}(\alpha) + \sum_{a,b} L_{a,b}(\alpha) 
\phi^a \lambda^b \right] \;\; , \label{ourresult}
\end{equation}
namely a number of $L_{1,b}$--terms and the leading term, $L_{2,2}$, in 
$O(\phi^2)$. All the calculated terms are given as analytic expressions. 
The first one, $L_{1,1}(\alpha) = 8 {\cal L}(\alpha) {\cal L}^\prime(\alpha)$,
yields the result (\ref{M7O11}) of Ivanov et~al. and
$L_{1,2}$ is the same as in the expression (\ref{M7O12}). A generalization
to the polydisperse case is given in \cite{BHML2003}. 

In the same year, Klapp and Patey \cite{KP2000} present results obtained from
two different integral equation approximation schemes that use either the MSA
or the RHNC as two-particle closures. The singlet density $\rho$ and the 
external  
field is related in formally exact ways either via the first BGY hierarchy 
equations 
\cite{HMcD90} to the pair correlation function $g(1,2)$ or via the 
Lovett-Mou-Buff-Wertheim equations \cite{LMB76,W76} to the direct correlation
function $c(1,2)$. The results turn out to
be strongly dependent on the form of the closure used. For example, the 
authors find in zero field
ferromagnetic behavior using the RHNC closure that is absent in case of the
MSA closure \cite{KP2000}.
  
Pshenichnikov and Mekhonoshin \cite{PM00a} present Monte--Carlo
results for a finite system giving $\chi$ as a function of $\chi_L$ for fixed
$\lambda = 1$, $3$, and $4$. They also calculate the equilibrium magnetization 
for a volume fraction $\phi = 0.226$ and $\lambda = 1$ and $4$. The results for 
$\lambda = 1$ can quite well described by the equation (\ref{M7O11korr}). The 
equilibrium magnetization for $\lambda = 4$ is however smaller than predicted 
by any theory for interacting particles and is explained as result of the
formation of clusters with a small total magnetic moment.

Ivanov and Kuznetsova \cite{IK01} perform a cluster expansion
calculating the $O(\phi^2 \lambda^2)$--term in $M/M_{sat}$. While their 
results are not given explicitly they seem to reproduce \cite{IK2001a}
the result of \cite{BHML2000}. The authors of \cite{IK01} suggest an extension 
of Eq.~(\ref{M7O11korr}) that reads
\begin{equation} \label{M7O22korr}
M = M_{\text{sat}} {\cal L} 
\left( \frac{m H_{\text{eff}}}{\kb T} \right) \; , \quad
H_{\text{eff}} = H + \frac{1}{3} M_{\text{sat}}{\cal L}(\alpha) + 
\frac{1}{144} M_{\text{sat}}^2 {\cal L}(\alpha) \frac{\dd }{\dd H} 
{\cal L}(\alpha) \; .
\end{equation}
A polydisperse generalization, replacing ${\cal L}$ by 
${\cal L}_{\text{poly}}$ is also given. 
The formula (\ref{M7O22korr}) produces sound magnetization curves for arbitrary 
$\phi , \lambda$ as in the case of Eq.~(\ref{M7O11korr}). When comparing the
results of \cite{IK01} with those of the cluster expansion \cite{BHML2000} one 
has to be aware that in \cite{IK01} 
some calculated terms that do not contribute to the initial susceptibility 
were discarded in order to be able to 
write the result in the form (\ref{M7O22korr}). Thus, when this equation is 
expanded in $\phi \lambda$ one does not recover the general
$O(\phi^2 \lambda^2)$--term of (\ref{ourresult}) while it is 
reproduced in the expression for the initial susceptibility.

Molecular dynamics simulations of monodisperse dipolar systems are performed 
in 2002 by Wang, Holm, and M{\"u}ller \cite{WHM2002} for several combinations of
$\phi$ and $\lambda$. They use a modified Stockmayer potential with a shifted
and truncated Lennard--Jones  potential. Comparing with the theory of Ivanov and
Kuznetsova \cite{IK01} they find good agreement for $\lambda \leq 2$ but 
deviations for higher $\lambda$, both for the magnetization curve and the 
initial susceptibility. The authors also demonstrate that $M$ and $\chi$ are not 
functions of the product $\phi \lambda$ alone as predicted in \cite{IK01} and
that the theory \cite{IK01} becomes less applicable for large $\lambda$ and 
small $\phi$. Furthermore they analyze the microstructure of the simulated fluid 
and determine the length distribution of chains. They show that the 
susceptibility can be better explained by replacing the Langevin
susceptibility in (\ref{CLEoep2}) by the susceptibility of a system of ideal
dipolar chains with a length distribution found in the simulations.
In 2004 Ivanov, Wang, and Holm \cite{IWH03} use a variant of the chain formation
model of \cite{ZI96} to obtain a length distribution independent of the simulation
results.
Already in 2003 Wang and Holm publish simulation results on
bidisperse dipolar systems \cite{WH2003}. For a system
containing particles with $\lambda = 1.3$ and $\lambda = 5.32$ and a total 
volume fraction of $\phi = 0.07$ they report that the theory of Ivanov and
Kuznetsova \cite{IK01} works well for volume fractions
$\phi_L \leq 0.02$ of the large particles. However, differences become apparent 
for larger fractions of the large
particles, say, $\phi_L = 0.05$  and even more so in the limiting case, 
$\phi_L = 0.07$, without any small ones, i.e., for a monodisperse fluid containing 
only large particles. 
Krist{\'o}f and Szalai \cite{KS03} perform in the same year Monte--Carlo 
simulations for two polydisperse ferrofluids with realistically chosen 
parameters. They find good agreement with the polydisperse version of 
(\ref{M7O11}).

Ra\c{s}a~et~al. \cite{RBPV02} measure in 2002 the initial susceptibility and
the magnetization of
ferrofluids at various stages of dilution. The susceptibility is found to be
best described by Ivanov's theory \cite{BI92} and the MSA \cite{ML90}.
\newpage
\section{Nonspherical potentials and higher-order magnetic moments}\label{SecV}

Real ferrofluid particles can be expected to have small deviations from a 
perfectly spherical shape. Such deviations 
affect not only the form of the nonmagnetic part of the interparticle 
potential. Also the magnetic interaction potential will be changed since 
the magnetic field of a nonspherical magnetized particle will contain also 
higher magnetic moments beyond the dipolar one. Both features will have some
influence on the susceptibility and the magnetization. The susceptibility of
systems of simple nonspherical dipolar particles and particles with dipolar 
{\it and\/} quadrupolar particles where also investigated on some occasions. 

A straightforward generalization of the system of DHS is the system of dipolar
hard ellipsoids. Joslin calculates \cite{J82} numerically the second 
dielectric virial coefficient for this system for different 
$\lambda$ and axis ratios. For spheres, this coefficient is positive, see 
Eq.~(\ref{CLEBJ80}). However, it depends strongly on the axis ratio: it grows 
with growing prolateness but it becomes negative already for only slightly 
oblate particles. Perera and Patey \cite{PP89} investigate this system 
by means of the HNC, finding the same trend.

Dimer ferrofluid particles were modeled by dipolar fused spheres
(dumbbells). A Monte--Carlo simulation of Stockmayer dimers can be found in a paper
of de Leeuw and Quirke \cite{LQ84}. $\chi$ is found to decrease with
growing sphere separation. Dimers of dipolar hard spheres
are simulated by Lomba, Lombardero, and Abascal \cite{LLA89}. They 
study the density dependence of the susceptibility for fixed dipolar strength
and find a decrease of $\chi$ at high densities due to the hindrance of particle
rotation.

Dipolar particles of spherical shape that carry, however, an additional 
quadrupolar 
moment are investigated with respect to their susceptibility by Patey, 
Levesque, and Weis \cite{CP79} 
and Carnie and Patey \cite{CP82} for different choices of the
quadrupolar tensor using integral theories. The susceptibility is shown to 
decrease with growing quadrupolar moment.

In view of the fact that the average shape of a ferrofluid particle is 
roughly spherical it seems to us that systems of identical 
nonspherical or multipolar particles that have been considered in the literature
are probably less applicable to describe real ferrofluids than systems of DHS or
Stockmayer particles. To improve upon monodisperse models of identical 
particles one would have to consider at
least binary mixtures of particles of different nature,
e.~g.\, of prolate and oblate ellipsoids or of monomers and dimers.
\newpage 
\section{Phase transitions} \label{SecVI}

Two different kinds of phase transitions are addressed in this
section. The first one is the condensation of magnetic particles into a denser
phase. The second one is the transition from an isotropic nonmagnetic phase 
to a state with spontaneous long--range magnetic order. 

There are other, less intensively investigated transitions in 
systems of dipolar particles such as fluid/solid transitions 
\cite{WL1993,GD96,GD01} or 
the demixing of particle species in bidisperse systems \cite{CKF91}. Another
example of a phase transition that we do not touch here is the emergence of a 
dipole 
glass state at low temperatures for systems of dipoles in a solid matrix, like 
frozen ferrofluids \cite{LNRR91,MLG94,AGP97}. The existence of such a state in 
liquid ferrofluids has been reported, e.~g., in \cite{MZL90}. But then it was shown 
that the observed temperature behavior of the susceptibility can be explained 
as an effect of the temperature dependence of the viscosity of the carrier 
liquid \cite{BR95}. A review of this subject for the electrical case can 
be found in \cite{VG90}.

\subsection{Dilute/dense phase transitions}

Ferrofluid particles possess an isotropic attractive potential: the outer tail 
of the van der Waals attraction that is not shielded by the steric or
electrostatic repulsion. This can result in the creation of small
clusters, as seen by Donselaar et al.\ in cryo--TEM images
\cite{DFBBHP99}.

When the attractive potential has a strength of about
$\kb T$ a dilute/dense phase transition should occur. It is possible to 
weaken the repulsive shielding of the ferrofluid particles enough to allow
such a reversible phase transition without reducing it so
much that an irreversible agglomeration takes place as observed in experiments 
with ionic ferrofluids \cite{BPSCM88,BPSCM90,MDCH95,CDC03}. This reversible 
transition is 
a kind of gas/liquid transition. It affects however only the magnetic subsystem 
of the ferrofluid -- the carrier system remains always liquid. We will use the 
term ''dilute/dense'' instead of  
''gas/liquid'' phase transition also when referring to the common model systems
without any carrier liquid to avoid a confusion of terms. The dense phase exists
in the form of (mesoscopic) droplets.

In a typical ferrofluid the isotropic interaction is rather weak. 
In Fig.~\ref{rosensweig2}, for example, the best fitting Lennard--Jones potential
has a depth of $v_0 = 0.16$. Since a dilute/dense transition does not occur in 
pure Lennard--Jones systems when  $v_0 < 0.76$ \cite{SG95b} one concludes 
that such a phase transition might possibly occur in ferrofluids with smaller 
$v_0$ only if the dipolar interactions have a net supporting, i.e., attractive 
effect.

A dilute/dense transition observed in experiment that might be caused mainly by
dipolar interactions was reported by Mamiya, Nakatani, and Furubayashi 
\cite{MNF00}. These authors investigate a ferrofluid consisting of iron--nitride 
in kerosene with $3 \lesssim \lambda \lesssim10$. They find a hysteresis in $\chi$ as
a function of temperature when $\phi$ is roughly between 0.002 and 0.06.
For larger $\phi$ the hysteresis vanishes. Furthermore, $\chi(T)$ is larger 
when cooled than when reheated in the hysteresis region. The authors 
explain this with the presence of closed rings or drops
of particles with mostly vanishing total magnetic moment in the low-temperature
phase.   
 
When studying phase transitions of dipolar hard spheres, the
dipolar coupling constant $\lambda \sim 1/T$ serves as the dimensionless inverse
temperature. Likewise, $\phi$ serves as dimensionless density. In the case of
Stockmayer particles both, $\lambda$ and $v_0$, provide an inverse temperature
scale and in general the latter is used. The other quantity or the temperature
independent quantity $\lambda/v_0$ serves as an additional parameter. 

Considering these main model potentials it is clear that a system of
Stockmayer particles will exhibit a dilute/dense phase transition below some
critical $T_c$ since it contains the Lennard--Jones potential as 
special case. The additional dipole--dipole interaction raises $T_c$ (see below)
an effect 
that cannot be explained trivially, since this interaction can be both 
repulsive and attractive and its orientational mean is zero. However, an 
averaging 
procedure as in (\ref{SYVeffdef}) that takes into account the fact that 
attracting configurations are preferred yields an attracting mean dipolar
contribution. 

In the case of dipolar hard spheres and also soft spheres it is even more 
unclear a 
priori whether such systems show a dilute/dense phase transition since here 
the dipolar interaction is the only one that could cause such a behavior. As 
we will discuss below, the question whether this phase transition exists or not  
in pure dipolar hard sphere systems is still being debated.

In systems of Stockmayer particles with small $v_0$ strong dipolar couplings seem
to be necessary in order to raise the critical temperature enough to observe a 
dilute/dense transition. For example, Frodl and Dietrich \cite{FD92} find   
using a density--functional theory that already a dipolar coupling constant of 
$\lambda \approx 1.5$ is necessary to lower the critical point to about 
$v_0 \approx 0.4$. According to Monte--Carlo calculations by Hendriks, van
Leeuwen et al.~\cite{SWHL89,LSH93,Lee94} between 1989 and 1994 it is
$\lambda \approx 2.35$ for $v_0 \approx 0.4$, and Monte--Carlo calculations of 
Stevens and Grest \cite{SG95} from 1995 give a very similar value here. That 
means that common ferrofluid 
particles with small $v_0$ will not show a dilute/dense phase transition within 
a reasonable temperature range except in systems where the dipolar interaction 
is the by far 
dominating  contribution. It may seem that in such a case the short--range 
attraction should play no more any role and that Stockmayer particles 
with high $\lambda$ should behave similar to dipolar hard spheres. However,  
as mentioned already this does not seem to be the case.

In the 1970s and 80s the dilute/dense phase transition of dipolar hard spheres
was studied by using mainly
analytical and semianalytical theories that were conceived to investigate the
equilibrium magnetization or initial susceptibility. A usual kind of 
$(V,T)$--phase diagram was found with a roughly parabolic coexistence curve 
and no phase transition above some critical $T_c$ (i.~e.~below some critical 
$\lambda_c$). However, the calculated critical values differed somewhat. 

Rushbrooke, Stell, and H{\o}ye \cite{RSH73} present in 1973 the coexistence
curve based on the MSA and a perturbation expansion ansatz by 
Stell, Rasaiah, and Narang \cite{SRN72}. An early Monte--Carlo
calculation is undertaken by Ng, Valleau, and Torrie in 1979 \cite{NVT79} using
only $N = 32$ particles.
Woodward and Nordholm \cite{WN84} use three similar expressions for the
free energy based on their effective potential (\ref{SYVeffdef}) to calculate 
the coexistence curve.
Kalikmanov et al.~\cite{BKF87,K92} use the same effective potential 
and replace it by a related Lennard--Jones potential to find the critical point.
Joslin and Goldman \cite{JG93} finally calculate in 1993 the critical point by
using second and third virial coefficients obtained from numerically calculated
cluster integrals. 
The approximate results for the critical point $(\phi_c, \lambda_c)$ are 
summarized in the following table:

\begin{center}
\begin{tabular}{r|c|c|c|c|c|c|c|c|}
Work & \cite{RSH73}(MSA) & \cite{RSH73}(SRN) & \cite{NVT79} & 
\cite{WN84}a & \cite{WN84}b & \cite{WN84}c & \cite{K92} & \cite{JG93} \\
\hline
$\phi_c$ & 0.056 & 0.083 & 0.15 & 0.17 & 0.14 & 0.13 & 0.16 & 0.028 \\
$\lambda_c$ & 4.4 & 3.6 & 4.0 & 2.3 & 2.9 & 2.9  & 1.5 & 3.8
\end{tabular}
\end{center}

Despite the different predictions for the critical point there was a general 
belief at that time
in the existence of a first--order dilute/dense phase transition. This changed 
when a couple of new Monte--Carlo simulations failed to detect such a transition.

Caillol \cite{Caillol93} performs a Monte--Carlo simulation with up to 
512 particles and does not find a phase transition for two coupling constants 
$\lambda = 4.5$ and $\lambda = 5.55$ and volume fractions 
$0.08 \frac{\pi}{6} < \phi < 0.38 \frac{\pi}{6}$.
Furthermore, Monte--Carlo calculations by Weis and Levesque \cite{WL1993,WL93}  
done for systems of similar size do not show a 
dilute/dense transition at low densities, $0.01 < \phi <0.16$, even for 
couplings as strong as $\lambda = 12.25$. Instead association of particles 
into chains is observed. 

Van Leeuwen and Smit \cite{LS93} consider a dipolar fluid with a modified
Lennard-Jones short--range potential of the form 
\begin{equation}
v^{SR}_{ij} = 4 v_0 \left[ \left( \frac{D}{r_{ij}} \right)^{12} - 
c \left( \frac{D}{r_{ij}} \right)^{6} \right] \;\; .
\end{equation}
The introduction of the parameter $c$ allows a smooth transition from 
dipolar soft spheres ($c = 0$) to Stockmayer particles ($c = 1$)
\cite{footnoteLS93}. For one fixed $v_0$ 
the authors do not find a dense--dilute phase transition if $c < 0.3$
and conclude that the dipolar soft and hard sphere fluid will not exhibit such a
transition. They observe chain formation instead.
 
Levesque and Weis \cite{LevesqueWeis94} investigate the system of DHS
with a dipolar coupling $\lambda = 12.25$ in an even larger range of volume
fractions, $0.005 \frac{\pi}{6} < \phi < 0.8 \frac{\pi}{6}$, and in addition the 
range $4 < \lambda < 12.25$ for $\phi = 0.01\frac{\pi}{6}$ and 
$\phi = 0.3\frac{\pi}{6}$. For $\lambda = 12.25$ they
find long chains when $\phi < 0.2 \frac{\pi}{6}$. For higher volume fractions the 
situation is less clear. Shorter chains can still be found applying an energetical 
criterion, but these chains are not directly visible in the configuration snapshots.
At $\phi = 0.6 \frac{\pi}{6}$ and higher they find a ferroelectric state (see below). 
They observe a continuous association of monomers into chains when increasing 
$\lambda$ for fixed volume fraction.

In 1994 and 1995 Stevens and Grest \cite{SG95,SG94} perform Monte--Carlo 
calculations for dipolar soft spheres both with and without applied magnetic 
field. They do not find a phase transition for the zero field case. Instead
they observe also in this 
system chain formation similar to that of DHS. However,
still in 1995 the same authors \cite{SG95b} report a dilute/dense phase
transition for Monte--Carlo simulated Stockmayer fluids where the critical 
temperature measured in $1/v_0$ and the critical density depend linearly on 
$\lambda/v_0$. The critical parameters agree well with those found by Hendriks, 
van Leeuwen et al.~\cite{SWHL89,LSH93,Lee94} who considered a smaller range of 
$\lambda/v_0$ already between 1989 and 1994.

The association of ferrofluid particles into chains for $\lambda \gg 1$ and low
densities is a natural consequence of the highly directional dipolar 
potential \cite{GP70,Jordan73}. The magnetic moments should be oriented in
these chains in the energetically most
favorable head--to--tail configuration with energy $-2 \lambda \kb T$ for a pair
of them. 
The numerical observation of chain formation instead of a homogeneous dilute/dense 
transition 
leads to a number of papers considering simple thermodynamic models of 
chains of dipolar hard spheres of similar particles. Small-angle neutron
scattering experiments \cite{RJR90} have provided experimental support for the 
existence of such chain structures. Recently, a direct visual confirmation of
the existence of chain structures in cryo--TEM images was achieved by Butter
et~al.~\cite{BBFVP03}.

Sear \cite{Sear96} argues that the chains should interact with 
each other only weakly and so he considers a model of only nearest--neighbor 
interacting particles, forming ideal chains. He compares with results of the 
simulations of \cite{LevesqueWeis94} and \cite{Caillol93}. Van Roij 
\cite{Roij96} investigates the competition between condensation due to isotropic
interactions and chain formation due to dipolar forces.

The most detailed model 
was proposed by Osipov, Teixeira, and Telo da Gama  
\cite{OTG96,TGO97} who also take into account chain--chain interaction. They 
compare with results from \cite{Caillol93} and \cite{WL93}. These authors find
semi-quantitative agreement concerning the critical points with the results of 
\cite{LS93} for Stockmayer--like potentials. They predict a dilute/dense 
phase transition to appear also for particles with weak isotropic attraction
(but not for DHS) where in both phases chaining should be observed. However, they
also note that finite-size effects may make it difficult to observe this
kind of transition in simulations. In 1999 Tavares, Weis, and Telo da Gama
\cite{TWG99} improve the theory and perform own Monte--Carlo calculations at 
low densities $\phi = 0.05 \frac{\pi}{6}$ and $5 \lesssim \lambda \lesssim 7.6$.
Also in 1999 Levin \cite{Levin99} uses the Debye--H{\"u}ckel--Bjerrum theory to
show that the clustering of particles into chains in the system of DHS should 
always reduce the density of free particles below the necessary density for a
phase transition.  
 
Klapp and Forstmann \cite{KF97a,KF97b} investigate Stockmayer and DHS 
fluids using the RHNC. There exists a parameter region of low temperatures where 
the RHNC fails to provide solutions. The boundary of this region is believed to be 
connected with phase transitions. At small densities the authors find growing 
fluctuations near this boundary that seem to represent the appearance of chains.
There are analogous hints for a dilute/dense phase transition in the case of
Stockmayer particles but not for DHS.    

By the end of the 90s there thus was a kind of consensus reached that DHS do not
show a dilute/dense phase transition because of the competing process of
chaining. But then
Shelley et~al.\ \cite{SPLW99} perform in 1999 Monte--Carlo simulations for 
dipolar hard dumbbells and spherocylinders. They find a dense/dilute phase 
coexistence even for almost spherical particles and observe, however, increasing 
sampling problems. They suggest that these sampling problems may have prevented 
the discovery of such a coexistence for DHS. 
Camp, Shelley, and Patey \cite{CSP00} therefore reinvestigate in 2000 the 
system of DHS with Monte--Carlo simulations. For $\lambda = 7.56$ they find 
indeed evidence for one or even two isotropic dense/dilute phase transitions. 
In the same year Camp and Patey \cite{CP00}  study the evolution of a system 
with $0.001 \frac{\pi}{6} \leq \phi \leq 0.6 \frac{\pi}{6}$. 
The simulation shows ring structures for the
smallest $\phi$, chain structures for intermediate densities 
($0.06 \frac{\pi}{6} < \phi < 0.35 \frac{\pi}{6}$), where the phase transitions 
reported in 
\cite{CSP00} take place, and a structureless dense liquid for still larger
$\phi$. 

Pshenichnikov and Mekhonoshin \cite{PM00,PM01} also perform Monte--Carlo 
simulations of DHS using a finite system. They find an ordinary 
dilute/dense transition with a critical point already at $\lambda_c \approx 3$ 
and $\phi_c = 0.034$. This result is at odds with previous ones possibly because 
the authors of \cite{PM00,PM01}
use instead of the generally preferred Ewald summation technique finite 
cylindrical and spherical systems with $1000$ particles. They remark that 
periodic 
boundary conditions may cause errors in the case of long--range forces. However, a 
finite system with $N = 2046$ particles was previously used by Levesque and 
Weis \cite{LevesqueWeis94} for a volume fraction near $\phi_c$ and much higher 
$\lambda$ without finding a phase separation.

Morozov and Shliomis \cite{MS02} study the intrachain correlations of 
chains of DHS for large $\lambda$ by performing a $1/\lambda$--expansion of the
needed statistical integrals. They find a persistence length of $\lambda/2$ 
for $\lambda \rightarrow \infty$. Considering nonideal chains they predict a
transition from chain to globule structures at $\lambda = 3.2$
 
As discussed above, chaining in monodisperse systems has a strong influence on 
the dilute/dense phase transition. When comparing with experiments done on real 
ferrofluids it should therefore also be noted that polydispersity affects
the lengths of chains as theoretically predicted by Kantorovich and Ivanov
\cite{KI02} and demonstrated in simulations by Wang and Holm
\cite{WH2003}.

The phase behavior of DHS and similar systems is further complicated when 
taking into account a nonvanishing magnetic field. 
Sano and Doi \cite{SD83} investigate the dilute--dense phase transition 
in the presence of a magnetic field in a randomly filled lattice of dipolar 
particles. In zero field they find a rather small critical 
$\lambda$ and a rather large critical concentration in this system. In general 
a magnetic field
makes a phase transition more likely and increases the density differences of
coexisting phases. But there is a $\lambda$--interval near the zero-field 
$\lambda_c$ where a phase transition occurs for small and large 
fields but not for intermediate fields. 

Stevens and Grest \cite{SG94,SG95,SG95b} who failed to find a dilute--dense 
transition for dipolar soft spheres in zero field demonstrate the existence of 
such a transition in Monte--Carlo simulations with applied fields. However, the 
two phases cannot be described as simple homogeneous gas and liquid phases 
since long chains are present
in both of them. For both dipolar soft spheres and Stockmayer particles they 
show that $\lambda_c$ is getting smaller in applied fields whereas the value
$\phi_c \approx 0.017$ remains mainly unchanged. In \cite{OTG96} however it is 
argued that this observed phase transition for soft spheres may be just an 
artifact of the insufficient system size. 

Kusalik \cite{Ku94a} does not 
find a phase separation in the system of dipolar soft spheres at high fields. 
However, his value of $\lambda \approx 3$ is smaller than those considered by 
Stevens and Grest who show that $\lambda_c > 5$ even for infinite fields.
The field dependence of $\lambda_c$ and $\phi_c$ in the case of Stockmayer
particles is also studied with Monte--Carlo techniques by Boda~et~al.\ 
\cite{BWLS96}. The authors find very similar results to those of 
Stevens and Grest. Szalai, Chan, and Tang \cite{SCT03} find also
qualitatively the same behavior when they study dipolar Yukawa particles using
the MSA.

Klapp and Forstmann \cite{KF99} investigate perfectly aligned DHS, 
i.~e., the infinite--field limit within the framework of the RHNC. They also do not 
find a dilute/dense phase transition up to $\lambda \approx 5$. 

Zubarev and Iskakova investigate an interacting variant \cite{ZI02,IZ02}
of their chain model that has already been used to calculate the equilibrium
magnetization in \cite{ZI96,AM99,IWH03} and show that in this model chain 
formation prevents a dilute/dense transition in the infinite field limit.   

Morozov and Shliomis \cite{MS02} find that nonideal chains gain stability
against the transition to a globular cluster in applied fields.

\subsection{Ferromagnetic phases}
\label{SEC:ferrophase}

Here we review papers that have addressed the question whether the
classical dipolar interaction between the nanoscale particles in a ferrofluid
suffices to induce long--range orientational order of the magnetic dipoles in a
liquid state without positional order, i.e., with homogeneously distributed 
particle positions.

Already the oldest model for the equilibrium magnetization of dipolar
interacting particles, namely the Weiss model allows ferromagnetic
solutions, i.~e., a spontaneously generated equilibrium magnetization 
$M \neq 0$ for $H = 0$: Eq.~(\ref{Weissformula}) reads in a needle--shaped probe 
geometry for which $H = H_e$ 
\begin{equation}
M = M_{sat} {\cal L} \left( \frac{m M}{3 \kb T} \right) \;\; 
\end{equation}
when $H = H_e = 0$. This equation has solutions with $M \neq 0$  when  
$M_{sat}m/3 \kb T = \chi_L = 8 \phi \lambda$ exceeds 3. Approaching this
critical value, $\chi_L = 3$, of the Langevin susceptibility the initial 
susceptibility (\ref{WOchiweiss}) of the Weiss model undergoes a divergence.
While the existence of a ferromagnetic 
state for such a rather large range, $\phi \lambda > 3/8$, of parameters
can be ruled out experimentally there are experimental hints for a 
ferromagnetic phase in the paper by Mamiya, 
Nakatani, and Furubayashi from 2000 \cite{MNF00} who find growing ferromagnetic
fluctuations in their iron--nitride ferrofluid at low temperatures for
$\phi \approx 0.15$. 

It should be noted that the above mentioned divergence of $\chi$
disappears when the next diagonal term in the
$(\phi,\lambda)$--expansion, i.e., the $\phi^2 \lambda^2$--term is included in 
the 
expression for the equilibrium magnetization \cite{BHML2000}. In this extended
Weiss model one has
\begin{equation}
M = M_{sat} \left[ \frac{1}{3} \alpha_{\text{eff}} - \frac{20}{9} \phi^2
\lambda^2 \alpha_{\text{eff}}- \;\; ... \;\; \alpha^3_{\text{eff}} 
+ O(\alpha^5_{\text{eff}}) \right] \;\; ,
\end{equation}
where $\alpha_{\text{eff}} = \frac{m}{\kb T} (H + \frac{M}{3})$ is the Weiss 
mean
field. In this case the additional $\phi^2 \lambda^2$--term prevents the
existence of ferromagnetic solutions for $H=0$.

Zhang and Widom \cite{ZW93} extend the Weiss mean field theory 
describing the local field as a stochastic variable and find that 
ferromagnetic phases do not exist for any value of $\lambda$ if the volume 
fraction $\phi$ is smaller than some critical value 0.295.  

At the end of the eighties theoretical investigations based on model
Hamiltonians \cite{PLP88} or model distribution functions \cite{BC89} lead to
the prediction of ferromagnetic states only for particles with sufficient 
prolateness or oblateness. Nematic ferro-- or antiferromagnetic phases 
were investigated with different methods in systems of dipolar ellipsoids and 
spherocylinders for example in \cite{PP89,GJ96,WR97,TOT98}.

But in 1992 MD calculations by Wei and Patey \cite{WP92a,WP92b} show the
existence of  ferromagnetic states in the system of dipolar soft spheres for
$v_0 = 1/1.35$ and $\lambda / v_0 = 9$ for volume fractions above 
$\phi \approx 0.65 \frac{\pi}{6}$. Since both, the soft and the hard sphere 
potential,  
are not only spherically symmetric but also purely repulsive both should behave 
very similar.  
Weis, Levesque, and Zarragoicoechea \cite{WLZ92} find indeed in 1992 in 
Monte--Carlo 
simulations ferromagnetic fluid and solid phases in the system of DHS in their
investigated range $0.8 \frac{\pi}{6} < \phi < 1.2 \frac{\pi}{6}$ and 
$\lambda = 6.25$. 
More results for this range are reported 1993 in \cite{WL1993}. Weis 
and Levesque \cite{WL93} find ferromagnetic states also at lower volume
fractions $\phi \approx 0.3 \frac{\pi}{6}$.    

Wei, Patey, and Perera \cite{WPP93} use density--functional methods to
investigate the transition to ferroelectric phases in  systems of dipolar hard
and soft spheres reproducing the numerical results qualitatively but not
quantitatively.
More systematic investigations of the phase diagrams of Stockmayer particles are
undertaken by Groh and Dietrich \cite{GD94a,ZW95R,GD94b} also using 
density functional methods. The critical temperatures are high. For 
$v_0 \approx 0.3$ 
for example they find a transition between a dilute, isotropic fluid and a 
dense, ferroelectric fluid already for $\lambda/v_0 = 4$ or 
$\lambda \approx 1.3$.

Zhang and Widom \cite{ZW94} investigate the phase diagram 
starting with the expression for the free energy of a van--der--Waals fluid 
with two additional terms modeling the magnetic properties. 
The authors find qualitatively similar phase diagrams: For high temperatures
there is a continuous transition from a dilute isotropic to a dense 
ferromagnetic liquid that becomes first order at smaller temperatures. If the 
short--range interaction energy is small enough compared to the dipolar
interaction energy then there exists a temperature interval where the
isotropic/ferromagnetic phase transition is preceded at smaller densities by a 
first--order transition between two isotropic phases having different densities.

Levesque and Weis \cite{LevesqueWeis94} find ferromagnetic states for
$\lambda = 12.25$ and $\phi > 0.6 \frac{\pi}{6}$. They however now argue
that the ferromagnetic phase reported in \cite{WL93} for smaller $\phi$ was 
probably just a slowly decaying nonequilibrium state.
Stevens and Grest \cite{SG95,SG95b} find spontaneously magnetized fluid phases 
of Stockmayer
particles for $\lambda = 4$ and $\phi > 0.47$. For soft spheres they find such
magnetic fluid phases for the investigated values $4 < \lambda < 9$
when $\phi$ is greater than a value that depends linearly on $\lambda$.
They see hints for a hysteresis in the magnetic--isotropic transition for $\lambda
= 16$.
 
Groh and Dietrich \cite{GD97a} study the phase transition properties of
nonspherical dipolar hard particles using a density functional ansatz. In the
special case of DHS they find no dilute/dense transition between isotropic
phases but an isotropic/ferromagnetic transition that becomes noncontinuous at
$\lambda > 1.33$ and $\phi \approx 0.2$. The authors admit that the stability of
the ferromagnetic phase is overestimated with respect to the simulations. 
That density functional methods are probably not useful for quantitative
investigations of the phase behavior is also pointed out by Osipov, Teixeira,
and Telo da Gama \cite{OTG97} and in a different context by Ivanov \cite{Iv03} 
and Morozov \cite{Mor03}.

Klapp and  Forstmann \cite{KF97a,KF97b} find fluctuations
at the high density boundary of the region of existence of isotropic RHNC
solutions for the DHS and Stockmayer systems. They also find ferromagnetic
solutions beyond that boundary.

Camp and Patey \cite{CP00} report in 2000 that they found in their Monte-Carlo
simulations ferromagnetic phases  at $\lambda \approx 7.5$ for 
$\phi > 0.5 \pi/6$.
A Monte--Carlo simulation of a DHS fluid with dipolar polydispersity, one with a 
bidisperse diameter distribution, and a binary mixture of dipolar and neutral 
hard spheres is undertaken by Cabral \cite{Cabral00} for $\phi = 0.8
\frac{\pi}{6}$ and $\lambda = 6.25$. He shows that the polydispersity reduces 
the ferromagnetic order. 

In 2003 Ivanov \cite{Iv03} investigates the BBGKY equation relating the
one-particle distribution function  $\rho({\bf r}, {\bf m})$ to the pair 
correlation function $g(1,2)$. He
employs for the latter the self-consistent expression used in density
functional approaches \cite{WPP93,GD94a,GD94b,KF99}. When expanding it up to
second order in the dipolar interaction the so approximated BBGKY equation shows
a bifurcation to a solution with spontaneously generated {\em ferrimagnetic} 
long-range order 
for $\chi_L > 12(\sqrt{5}-2) \simeq 2.83$, i.e., slightly below the
bifurcation threshold for ferromagnetism in the Weiss model. Ivanov 
\cite{Iv03} argues that this peculiar ferrimagnetic behavior is an artifact that
is ultimately due to the mean-field
character of the density functional ansatz for the pair correlation function.

Then in 2003 Morozov \cite{Mor03} investigates the bifurcation properties of 
the so-called Lovett-Mou-Buff-Gubbins integral equation \cite{LMB76,Gu80} that 
relates the one-particle probability distribution function to the direct 
correlation function $c(1,2)$ and that is equivalent to the above mentioned 
first BBGKY equation. He establishes criteria for the bifurcation of solutions 
with long--range
magnetic order in ellipsoidal and spherical sample shapes in vacuum and compares 
with mean--field predictions, density function theory, and MSA results. Using a
generalized MSA type expression for $c(1,2)$ he finds
that the strength of short--range correlations plays a decisive role for the 
appearance of spontaneous long--range magnetic order --- the 
susceptibility diverges when the former exceed a critical strength. More
qualitative arguments then show that the short--range correlations are most
likely to be antiferromagnetic. He also shows that approximating $c(1,2)$ by the
two lowest orders in its diagram expansion always gives rise to liquid
ferromagnetic solutions.

At the end of this section on phase transitions in ferrofluids we should like to
quote Teixeira, Tavares and Telo da Gama \cite{TTT00}. They have presented 
in 2000 a fairly extensive review on the effect of dipolar forces on the 
structure and
thermodynamics of classical fluids that covers many aspects of the problems 
related to gas/liquid condensation phase transitions in various 
dipolar systems. Since their summary of the state of research on these problems 
still gives in our opinion an
adequate picture of today's state of the art --- also concerning the problem of
liquid ferromagnetism --- we give two quotes\cite{TTT00}:
{\em "Although widely studied in statistical mechanics, the phase diagrams of dipolar
fluids in general, and of  strongly dipolar fluids in particular, have remained 
largely uncharted"} and {\em "In conclusion, it is fair to say that a theory is not 
yet available that is capable of describing dipolar fluids over the whole range 
of densities and dipole strengths. Moreover, the mechanisms driving the phase 
transitions (as well as the location of the phase boundaries) remain unclear."}

\clearpage
\section{Conclusion} \label{SecVII}

Ferrofluids, i.e., suspensions of magnetic nano--particles in liquid carriers
are not only of technological interest but they continue to be also the object 
of many basic research projects. Among those has the effect of dipolar
interactions that are important in ferrofluids attracted considerable interest
for two reasons: First, dipolar forces are long--range so that, e.g., the 
equilibrium magnetization (or polarization) is geometry dependent.
Second, dipolar forces are attractive or repulsive depending on the relative 
orientation of the particles and their dipole moments with the orientational 
mean being zero. A realistic theoretical description of the colloidal suspension
will also take into account other interactions between the  particles and in
addition the 
dispersion of particle diameters and magnetic moments in ferrofluids that is 
absent in molecular systems. 
The dipolar interaction affects naturally the equilibrium magnetization, but
also the phase transition behavior. These two subjects were reviewed here.

The question of how dipolar interactions influence the initial susceptibility 
$\chi$ 
was studied for electrically polar systems long before the advent of 
ferrofluids. The oldest models, the Weiss and Onsager models, to capture this
influence with mean--field approaches were proposed already in the first half of
the 20th century. 
Only after 1970 other theoretical methods were used intensively to investigate 
$\chi$ by cluster expansions, integral equations, and Monte--Carlo and 
molecular dynamics simulations.
Here mainly dipolar hard spheres and Stockmayer particles were investigated.
While both represent only poorly polar molecules like water they are
much better approximations to the nanoscale magnetic particles of ferrofluids.
In fact the latter seem to be very good realizations of these two model systems.
Many of the earlier analytical and semi--analytical theories that originally 
were conceived for molecular systems were also extended to account for 
polydispersity in ferrofluids.
  
When comparing cluster expansion methods and integral equation methods with 
numerical simulations it seems that the RHNC theory and the simpler perturbation
theory by Tani~et~al.~\cite{THB83} agree best with the Monte--Carlo results for 
$\chi$, at least for
relatively high densities. The second theory has also been applied to explain
experimental data for a highly concentrated ferrofluid, as has a related 
cluster expansion result by Ivanov and Kuznetsova \cite{IK01}.

With the synthetization of ferrofluids, i.e., dipolar systems in which 
saturation could easily be reached not only $\chi$ but also the full 
magnetization curve $M(H)$ became of interest. Here, the 
main theoretical results date from 1985 onwards. The proposed theories were 
mostly
extensions of those already used to describe $\chi$. For
example, the one by Ivanov and Kuznetsova \cite{IK01} compares well with 
experimental data. However, recent numerical simulations by Wang et al.
\cite{WHM2002,WH2003} for $\chi$ and $M(H)$ are not well described by this theory 
when the volume fraction of the magnetic particles is small and the dipolar 
coupling is strong: they are 
better explained by a model of small non--interacting chains of different
lengths.

An ever increasing analytical, numerical, and experimental research activity
has been devoted to phase transition phenomena in dipolar systems, namely, 
({\em i}) the separation into dilute and dense phases and ({\em ii}) the
appearance of spontaneous (ferromagnetic) long--range order in zero magnetic
field. Both phenomena 
refer in the case of ferrofluids solely to the subsystem of the magnetic 
particles without positional long--range order. 

Dilute/dense transitions occur in Lennard--Jones systems and, more generally,
whenever the attractive part of the interaction is sufficiently strong. In fact
simulations have shown that switching on an additional dipolar interaction in
such systems favors the transition and increases the critical temperature. 
However, the question whether such a phase transition occurs also for purely
dipolar hard spheres without any isotropic attraction is more difficult to 
answer. The fact that such phase separation has been observed experimentally in
real ferrofluids does not help much to solve the above theoretical problem when
(strong) isotropic interactions cannot be excluded to be present in the
ferrofluid.

In the 1970s and 1980s a variety of semi--analytical theories was applied to
investigate the dilute/dense phase transition behavior of dipolar hard spheres.
They all found a usual phase diagram with a
roughly parabolic coexistence curve. However, there was no agreement about
the location of the critical point.
But Monte--Carlo simulations in the 90s did not find such phase
transitions for dipolar hard spheres. For strong enough dipolar
couplings or, equivalently, low enough temperatures a formation of 
head--to--tail dipolar chains was observed instead. They are caused by the
highly directional character of the 
dipolar pair potential. Thus, new models were developed to explain the 
observed behavior by means of polymer theory. According to these models the 
assembly of chains prevent a usual dilute/dense transition or mask it and 
make its observation in simulations difficult due to finite--size effects.

Another turning point was reached in 2000 when Camp et al. \cite{CSP00}
found evidence for one or even two phase transitions between chain--dominated
states in Monte--Carlo simulations.  Pshenichnikov and Mekhonoshin 
\cite{PM00,PM01} even observed the common dilute/dense transition already at 
quite low
dipolar couplings where chain formation is not very effective. They
explain the differences to other studies with their choice of a finite system
instead of periodic boundary conditions. The differences between 
both simulation methods remain to be investigated further.

The phase transition behavior is complicated further by the presence of a 
magnetic field.
In the case of Stockmayer particles the field supports the phase
separation, i.~e., it enlarges the coexistence region. The question whether 
a field can 
trigger such a phase separation also for dipolar hard spheres is not yet 
settled.

In comparison to the question of a dilute/dense transition the research results 
concerning the possibility of having a 
phase with spontaneous (ferromagnetic) long--range order in zero magnetic field
seem to be more coherent --- albeit only on first sight:
Experimental hints for such a phase were found by Mamiya et al. \cite{MNF00} and
the simulations performed since the 1990s generally agree that there exist 
ferromagnetic phases for dipolar hard spheres and similar systems in regions of
high densities and strong couplings. This behavior was also found
in density functional theories although they seem to be less suited to
explain the phase behavior and the results differ quantitatively.
The conditions for the appearance of spontaneous long--range magnetic order 
were recently investigated  by Morozov \cite{Mor03} and Ivanov \cite{Iv03} using 
integral equation methods. They conclude that mean-field type
approximations that are also the core of density functional approaches tend to
generate artificially long--range magnetic order.

So, all in all we think that a lot of questions/problems related to the
equilibrium behavior of ferrofluids and dipolarly interacting particles remain 
to be addressed.

\section*{Acknowledgments}   
We thank K. Morozov for helpful discussions and for critically reading the 
manuscript. This work was supported by the DFG (SFB 277) and by INTAS (Ref. Nr.
03-51-6064).


\clearpage
\clearpage
\clearpage
\clearpage

\begin{thebibliography}{99}
\bibitem{error-BM93-ASADOL93}
In \cite{BM93} the last term in Eq.~(44) must read 
$32( \sinh \alpha /\alpha - 2 \cosh \alpha /\alpha^2 + 
2 \sinh \alpha/\alpha^3)^2$. In \cite{ASADOL93} the last term in Eq.~(12) 
must read $3 i_2^2(\alpha)/\alpha^2$.
\bibitem{footnoteLS93}
It should be noted however that the potential for $c \neq 0$ can still be 
written as Stockmayer potential with different $v_0$ and $D$.
\end{thebibliography}
\end{document}